\documentclass[showpacs,amssymb,preprint,preprintnumbers]{revtex4}
\usepackage{amsmath}
\usepackage{xcolor}   

\begin{document}

\title{Quantum Noise of Gravitons and Stochastic Force \\ on Geodesic Separation}
\author{H. T. Cho}
\email{htcho@mail.tku.edu.tw }
\affiliation{Department of Physics, Tamkang University,
Tamsui, New Taipei City, Taiwan, R.O.C.}
\author{B. L. Hu}
\email{blhu@umd.edu}
\affiliation{Maryland Center for Fundamental Physics and Joint Quantum Institute,\\ University of
Maryland, College Park, Maryland 20742-4111 U.S.A.}

\date{v.1 Dec. 15, 2021, v.2 Feb. 20, 2022}

\begin{abstract}
In this work we consider the effects of gravitons and their fluctuations on the dynamics of two masses using the Feynman-Vernon influence functional formalism, applied to nonequilibrium quantum field theory and semiclassical stochastic gravity earlier by Calzetta, Hu and Verdaguer \cite{CalHu94,CalHu08,HuVer20}, and most recently,  to this problem by Parikh, Wilczek and Zahariade \cite{PWZ20,PWZ21a,PWZ21b}. The Hadamard function of the gravitons yields the noise kernel acting as a stochastic tensorial force in a Langevin equation governing the motion of the separation of the two masses. The fluctuations of the separation due to the graviton noise are then solved for various quantum states including the Minkowski vacuum, thermal, coherent and squeezed states. The previous considerations of Parikh et al. are only for some selected modes of the graviton, while in this work we have included all graviton modes and polarizations. We comment on the possibility of detecting these fluctuations in primordial gravitons using  interferometors with long baselines in deep space experiments. 
\end{abstract}
\maketitle 

\section{Introduction}

After the first observation of gravitational waves by LIGO in 2015 \cite{LIGO}, the efforts to understand the properties of these waves have been thriving. This and the subsequent observations have confirmed the classcial theory of Einstein general relativity in a convincing way. However, is it possible that these interferometry observations could reveal the quantum nature of gravity? Recently, a proposal has been put forth by Parikh, Wilczek, and Zahariade (PWZ)\cite{PWZ21a,PWZ21b} on this possibility by trying to measure the quantum noise effects due to gravitons. This has aroused much interest in the detection of gravitons by laboratory experiments or other means (e.g., \cite{Kanno2}).

The detection of individual graviton was deemed to be nearly impossible \cite{Dyson}. Instead, the detection proposal of PWZ focuses on the effects of the quantum noise of gravitons on classical particles or masses in the same vein as the quantum Brownian motion (QBM) paradigm \cite{Schwinger61,FeyVer63,CalLeg83,HPZ}. For gravitons we need two masses since gravitational effects show up only in geodesic separations and their fluctuations.   In the Feynman-Vernon \cite{FeyVer63}  influcence functional treatment of QBM the quantum degrees of freedom of the graviton manifest in both the dissipative and noise effects on the particles. The application of this influence functional formalism to open quantum systems \cite{qos} and the Schwinger-Keldysh \cite{Schwinger61,Keldysh64} or closed-time-path integral \cite{Chou85,CalHu87,CalHu88} or `in-in' \cite{Wein05} method has a long history (see, e.g, \cite{CalHu08,JH0,JH1,QRad,GalHu05,GHL06,StoGraLivRev, HuVer20} and references therein).  A  synopsis of this approach as applied to two related set of problems earlier  is given in the next section.

In the  considerations of PWZ \cite{PWZ21a,PWZ21b}, to simplify the formalism, the authors have computed only a selected set of modes with certain polarizations.  Here,  in this paper, we work with a more general framework where all graviton modes and polarizations are taken into account. As the graviton field can be viewed as two massless minimally-coupled scalar fields corresponding two different polarizations \cite{FP77}, these scalar degrees of freedom can be integrated over via the closed-time-path integrals. When the Feynman-Vernon formalism is applied to a Gaussian system, quantum noise  can be represented by a classical stochastic force with the correlation function given by the noise kernel.  Variation of this stochastic influence action leads to a Langevin type of equation with this stochastic force as source.  In a nutshell, the noise and the dissipation kernels in the influence functional produce  respectively the noise and the dissipative effects of the gravitons on the geodesic separation between particles or masses.  

Gravitons may exist in different quantum states, the immediately relatable one is of course the Minkowski vacuum. On the other hand, like the cosmic microwave background, the primordial gravitons in the present universe could be in a very low temperature thermal state. While in the early universe  the gravitons could be in a very high temperature state, if they had already come to thermal equilibrium. Hence, it is necessary to investigate both the low temperature and the high temperature limits of both the noise kernels and the  separation fluctuations in these cases. From the theory of cosmological particle creation it is known that   particles created by the expansion of the universe exist in a squeezed vacuum state \cite{GS90}, so would primordial gravitons. Thus it is necessary to examine the effects of gravitons in a squeezed quantum state on  the separation fluctuations of particles or masses.

The plan of the paper is as follows. In Section II, in view of the increased attention attracted to the recent experimental proposals to test the quantum nature of gravitation,  it is perhaps helpful to point out some common confusions about quantum gravity and graviton physics.  A synopsis of the open quantum system approach  is  given as applied to two related subjects studied before -- radiation reaction in moving masses and charges and semiclassical stochastic gravity -- with the hope that  readers can gain a broader perspective and learn from what had been done earlier with the same methodology.  Readers familiar with these topics and approaches please skip over this section and go directly to Section III,  where the stochastic effective action, with the dissipation kernel and the stochastic tensor force is derived. From this we derive the Langevin equation of motion for the geodesic separation and solve for the separation fluctuations. In Section IV, the noise kernels in four different quantum states-- the vacuum, the thermal, the coherent and the squeezed states -- of the gravitons are worked out explicitly. Then in Section V, the geodesic separation and the corresponding fluctuations are obtained. These results enables us to make estimates on the detectability of graviton fluctuations with the gravitons in different quantum states. We draw our conclusions in Section VI.  In the Appendix, the properties of the polarization tensors and their integrations over solid angles are discussed in some detail as they are used in various stages of the calculation.

\section{Graviton Physics and  Open Systems Approach -- a synopsis }

 This largely pedagogical section is aimed at serving two clarification purposes: a) to avert common confusions about quantum gravity and graviton physics, and b) to provide a synopsis of the open quantum systems approach using  two  fully studied earlier programs (with explicit pointers to equations therefrom, for direct and easier comparson), that of semiclassical and stochastic gravity \cite{StoGraLivRev,HuVer20} and  radiation reaction of moving charges or masses \cite{JH0,JH1,Barack}.

\subsection{Graviton physics is quantized perturbative gravity, not quantum gravity proper}

Laboratory experiments of analogy gravity \cite{AnalogG} have proven very fruitful in the past two decades. With the rise of a new subfield known as gravitational quantum physics \cite{GQP} proposals of  tabletop experiments for `quantum gravity' have also appeared in recent years \cite{Carney} and attracted immediate attention. While the proposed experiments have their own merits, it is perhaps necessary to  add a quick reminder that in actuality they  \cite{Bose17} are not about quantum gravity proper at the Planck scale  (see point 3 below), nor about the quantum nature of gravity \cite{Vedral17} (see point 4 below).  Now that the term `quantum gravity' is used by a broader range of researchers outside of the  gravitation/cosmology and particles/fields communities, certainly a welcoming development,  it might be helpful to make precise the physical meanings of the key terminology used.  For example, even a simple claim like, ``graviton physics is quantum gravity" without qualification can be misleading.  The strict answer would be no, because gravitons may not be the fundamental constituents of spacetime, as fundamental strings are meant to be,  citing one example of many candidate theories of quantum gravity proper.    Gravitons are quantized weak perturbations of spacetime, which are not the same as the basic constituents of spacetime, just like phonons are fundamentally different from atoms (in fact, phonons no longer exist at the atomic scale).   Superstring theory,  loop quantum gravity,  causal dynamical triangulation, causal sets  are better known examples of theories of quantum gravity proper,  which target fundamental spacetime structures above the Planck energy ($10^{19}$ GeV).  One could view all theories derived from quantum gravity proper at energy scales below the Planck  energy as effective field theories (EFT), ranging from semiclassical stochastic gravity to general relativity, but the perturbative nature is the more definitive feature in graviton physics \footnote{This includes theories valid at immediately below the Planck energy, such as a) dilatonic gravity with the 3-indexed antisymmetric tensor field, which had been proven to be a low energy (sub-Planckian) limit of string theory, and, on the same footing, b) semiclassical gravity, where the Einstein-Hilbert action is augmented by three additional terms ( $\Box R, Ricci^2, Weyl^2$)  resulting from the regularization of the stress energy tensor of quantum matter fields.  EFT also has been referred to at today’s low energy by authors like Donoghue and Burgess.   Therefore mentioning EFT is not sufficient to signify the specific nature of gravitons, namely, the quantized weak gravitational fields from the perturbations of an assumed background spacetime. The concept of graviton is embedded in the perturbative treatment.}.   There could be gravitons at all   scales greater than the Planck length scale ($10^{-33}cm$)  because spacetime as manifold now exists (whether spacetime manifold emerges from the interaction of its basic constituents is a crucial challenge of any proposed quantum gravity theory) from which one can consider weak perturbations off this classical entity and quantize them. 

In this light,  the answer to the  question we posed would be yes, if one adds the term `perturbative' to quantum gravity,  because gravitons refer  to the quantized dynamical degrees of freedom of perturbative (linearized) gravity,  which can be seen at today's low energy \footnote{We use ``perturbative quantum gravity” in the sense of Feynman beginning in the 50s and Weinberg in the 60s, which are primarily followed by the particle physics community, namely, in the context when graviton as spin-two particles can be defined.  There, the background spacetime (e.g., Minkowski, Schwarzschild or Robertson-Walker) dynamics is derived from the Einstein-Hilbert action. As is well-known,  perturbative quantum gravity is non-renormalizable. That has spurned heroic efforts, e.g., by Tsamis and Woodard, in the calculation of two-loop graviton interactions. The theoretical framework of such activities is perturbative quantum gravity.  On the other hand, dissatisfaction in its nonrenormalizable nature and background spacetime dependence (mostly in the general relativity community) prompted the development of loop quantum gravity, causal dynamical triangulation, causal sets, etc, in their pursuit of nonperturbative theories of quantum gravity.}.  If detected, we can indeed confirm the quantum nature of perturbative gravity,  as a very low energy effective theory of quantum gravity proper,  or as the quantized collective excitation modes of spacetime, similar to phonons (but not atoms).  For this purpose we give some background description below  to  define more clearly the contexts of terms used and their different focuses.  Hopefully this could help mitigate possible confusions from cross-talks over different intended purposes.

\subsubsection{Classical Gravity: General Relativity. Perturbative Gravity: Gravitational Waves}

Let us agree that when only `gravity' is mentioned it is taken to be classical gravity.  Classical gravity is described with very high accuracy by Einstein's general relativity (GR) theory.  One can distinguish two domains, weak field, such as experiments on earth or in the solar system would fall under, from strong field, such as processes near black holes or neutron stars,  depending on their masses and the proximity of measured events, and in the early universe.

Classical \textit{ `perturbative gravity'}  refers to small perturbations off of a classical gravitational background spacetime.  In the earth's environments, the background spacetime is Minkowski space. While the background spacetime could be strongly curved, perturbative treatments can only consider small amplitude deviations which are weak by proportion. Gravitational waves  are usually treated as perturbations  whose wavelengths can span from the very long of astrophysical or cosmological scales to the very short. Note the crucial difference between \textit{perturbations} and \textit{fluctuations}, the former being a deterministic variable, referring to the small amplitude deviations from the background spacetime, while the latter is a stochastic variable,  referring to the noise. Fluctuations in the gravitons constitute a noise of gravitational origin and of a quantum nature. Such effects at low energies are the target of the present investigation  \cite{PWZ21a,PWZ21b}, not Planck scale physics.  Fluctuations of quantum matter field can also induce metric fluctuations.  The metric fluctuations are governed by the Einstein-Langevin equation of semiclassical stochastic gravity theory \cite{HuVer20}. At the Planck scale they make up the spacetime foams \cite{foam} where topology changes can also enter.  

 \subsubsection{Quantum Gravity at Planck Energy: Theories for the microscopic constituents of spacetime} 

Quantum gravity (QG) proper refers to theories of the basic constituents of spacetime at above the Planckian energy scale, such as string theory, loop quantum gravity, spin network, causal dynamical triangulation, asymptotic safety, causal sets, group field theory, spacetime foams, etc \cite{Oriti}. Because of the huge  energy scale discrepancy between the Planck scale and the scale of a Earthbound or space laboratory, many such experimental proposals to test Planck scale quantum gravity need to rely on \textit{indirect} implications to high-energy particle phenomenology \cite{QGPhen,Sabine} or, at a lower energy range, analog gravity experiments \cite{AnalogG}. Of the latter kind, many atomic-molecular-fluid, condensed matter-BEC or electro-optical-mechanical experiments can indeed skillfully use analogs to seek indirect implications of quantum gravity. But in terms of \textit{direct} observations, or drawing direct implications, tabletop experiments can only probe weak-field perturbative gravity, nevermind their quantum gravity labels.

 \subsubsection{Gravitons can exist at today's low energy. They carry the quantum nature of  perturbative gravity.}

Gravitons are the quantized propagating degrees of freedom of weak perturbations off of  a background metric,   such as the Minkowski spacetime for experiments in the Earth's environment.
Graviton as a \textit{spin 2 particle} refers to the high frequency components of weak gravitational  perturbations under certain averages (like the Brill-Hartle \cite{BriHar}-Isaacson \cite{Isaacson} average), or in the ray representation under the eikonal approximation.

% As such, beware of the shortcomings when relying on gravitons  (lacking phase information)  to address  fully the quantum information issues of perturbative quantum gravity:  quantum coherence,  correlations and entanglement between gravitons need be considered as a whole.

 Gravitons can exist at today's very low energy but there is no necessary relation to the basic constituents of spacetime at the Planck energy. One does not need any deeper level theory for their description. Einstein's general relativity theory plus second quantization on weak linear perturbations will do. Note that any linearized degree of freedom in classical systems can be quantized, irrespective of whether it is fundamental or collective. The latter is in abundance in condensed matter physics (e.g., phonons, rotons, plasmons, and many other entities with -on endings).
Seeing the quantum nature of the gravitational field at today's low energy, such as proving the existence of gravitons \cite{Dyson,PWZ21a,PWZ21b},  is certainly of fundamental value, but there is no necessary relation to quantum gravity proper as defined above. Gravitons are the quantized collective excitation modes of spacetime, not the basic building blocks of spacetime. Graviton's existence is predicated on the emergence of spacetime while the building blocks (such as strings, loops, causets, etc) are the progenitors of spacetime structure.

These explanations are enough for  discerning graviton physics from quantum gravity.  Since there is increased enthusiasm  in using quantum entanglement of masses to show the quantum nature of gravity we should add one more noteworthy aspect. 

\subsubsection{Pure gauge says nothing about the quantum nature of gravity}
%Only the radiative degrees of freedom can testify to the quantum nature of (perturbative) gravity} %.

 A more subtle yet important misconception is attributing quantumness to the pure gauge degrees of freedom (Newton or Coulomb forces) while only the dynamical degrees of freedom (graviton or photon) are the true signifiers of the quantum nature of a theory.  Experiments measuring the entanglement between two quantum objects \cite{Bose17, Vedral17}, albeit through Newtonian gravitational interactions, only expresses the quantum nature of these objects, not of gravity.
This critique is raised in \cite{AnHucr}.  A more detailed explanation can be found in \cite{2GravCat}. Linearization of the GR field equations around Minkowski spacetime leads to an action similar to the EM one. The true dofs of the linearized perturbations are their transverse-traceless (TT) components, i.e., gravitational waves. Quantization of the short-wavelength TT perturbations  gives rise to gravitons. Only in the detection of gravitons, just like photons in QED, can one make claim to the quantum nature of gravity, but keep in mind their fundamental differences from quantum gravity proper as defined above.

\subsection{Open quantum systems and influence functional approach to graviton physics}

To help interested novices to fast-pace into the nonequilibrium dynamics of graviton physics we provide a synopsis of the open quantum systems concepts and the influence functional method, using two well-studied earlier examples:  stochastic gravity and   self-force. We first describe the theoretical framework and then point to the relevant equations in these two problems as illustrations.

The theoretical framework  used here and  in \cite{PWZ21a,PWZ21b} invokes techniques in quantum field theory known as the  Schwinger-Keldysh \cite{Schwinger61,Keldysh64},  close-time-path (CTP) \cite{Chou85,CalHu87,CalHu88} or `in-in' \cite{Wein05} formalism,  and concepts in open quantum systems \cite{qos},  a branch of nonequilibrium statistical mechanics \cite{CalHu08},   exemplified and embodied in the Feynman-Vernon influence functional formalism \cite{FeyVer63}.  Readers can find the different sources from the references given in this paper and in those cited in the Introduction.  However, a faster way to see their architectural structure perhaps is to compare our present problem with  two research programs, spanning the past three decades, using these methods. The first program (A) encompasses mostly cosmological and black hole backreaction problems, the second  (B) encompasses  gravitational radiation reaction problems.   The  problem we are studying here (C) falls under the second program.

A) Semiclassical and stochastic gravity \cite{StoGraLivRev,HuVer20}: Effects of quantum matter fields on classical spacetime dynamics. 

B)  Radiation reaction on   moving charges \cite{JH0,JH1,QRad} or masses \cite{GalHu05,GHL06} via stochastic effective field theory. 

C) Present problem: Effects of gravitons and their noises on the geodesic separation between two masses.

\subsubsection{Dynamical variables of the system  under the influence of an environmental field}

The first step, and sometimes the most tantalizing step, is to identify and ensure that certain variables can be treated as the system while other variables can be treated as its environment. The easy cases are those where a  discrepancy parameter can be clearly identified, like slow vs fast variables,  heavy versus light masses, high  vs low frequencies, etc.  For the two classes of problems we have mentioned in A) and B), the  gravitational sector being the system is clearly separated from the quantum matter field acting as its environment.  

For A) the system can be  {\it the weak linear metric perturbations $h_{\mu\nu}$} off of a curved  background  spacetime with metric  $g^0_{\mu\nu}$.  Assuming, for simplicity, that the background spacetime is a spatially-flat  FLRW (Friedmann-Lamaitre Robertson-Walker) universe with line element $ds^2 = -dt^2 + a(t)^2 (dx^2+dy^2+dz^2)$ and scale factor $a(t)$, then the small anisotropy $\beta_{ij}$ or  inhomogeneities $h_{\mu\nu}$ are the weak  metric perturbations which make up the system. The environment can be a quantum matter field $\Phi ({\bf x},t)$. We are interested in how the quantum matter field and its fluctuations back-react on the spacetime dynamics. 
  
For B) the system is {\it the charge or the mass, the dynamical variable being its trajectory $z(\tau)$}, with $\tau$ its proper time. We are interested in how a quantum field and its fluctuations back-react to determine the trajectories of the charge or the mass, in a self-consistent manner.  

For C), our problem at hand, the system refers to the two masses which serve as the detector, {\it the dynamical variable being their geodesic separation $ z^\mu$}. The gravitons and their noises are considered as its environment. 

Some  confusion in the role of the gravitons may show up between C) versus A) \&  B), but they can be distinguished in two easy ways:   a) for all three cases, the system is classical, while its environment is quantum.  Thus overall this is a semiclassical formulation.  b) Since each of the two polarizations of the graviton behaves like a massless minimally coupled scalar field, we can treat it in this manner.  Thus for all three cases the environment  is a quantum scalar field, albeit   a massless conformally coupled scalar field is often invoked in A) \& B) to make use of the convenience that there is no particle creation in a conformally-flat background spacetime such as FLRW, thus showing the net effects of the anisotropy and inhomogeneities.

\subsubsection{The environmental variables being integrated over: mean value  of the stress energy tensor for  semiclassical treatment}

This is usually done  by taking the vacuum expectation value  (vev) of the stress energy tensor (SET)   operator  of the dynamical variables of the environment influencing the system. For all three cases the environment is a quantum field (for Case C   the graviton field, which   can be represented by two massless minimally coupled scalar field).   For A) the vev of SET of a quantum field acts as the source driving the semiclassical Einstein equation. The theory thus formulated is called {\it semiclassical gravity}.  
One can use an effective action method, such as the approach taken in \cite{Hartle77}, with an interaction action of the form Eq. (2.5) of \cite{FHH79},
$ S_{int}(g;\Phi]= \int d^4x \sqrt {-g}  h_{\mu\nu}T^{\mu\nu}$ where  $h_{\mu\nu}$,  the metric perturbations, are the system variables, $g$ is the determinant of the background metric $g_{\mu\nu}$, and $T^{\mu\nu}$ is the stress tensor of the quantum matter field, acting as the  environment. See  \cite{HarHu79} for $h_{\mu\nu}$ representing weak anisotropy,   and  \cite{CamVer94} for weak inhomogeneities.   

Both Cases A \& B assume a quantum scalar  matter field as the environment.   In Case B, for moving charges and masses, the interaction action  is 
$ S_{int}(z;\Phi)= \int d^4x \sqrt {-g} j(x; z)	\Phi(x)$ where the current density $j(x; z)$	is some functional of the
worldline coordinates $z$ describing the motion of a charge or a  mass.  For a scalar charge it is given by Eq. (2.5) of \cite{GalHu05}. For an electric charge in an electromagnetic field, the current is a vector current $j^\mu(x; z)$,  and $\Phi$ is replaced by the vector potential $A_\mu$, see Eq. (3.2) of \cite{GHL06}.   For  a small point mass $m_0$ moving with 4-velocity $u^\mu$ through a gravitational field $h_{\mu\nu}$ of weak  metric perturbations  in a curved  background  spacetime with metric  $g^0_{\mu\nu}$,   the interaction is of the form given above, proportional to $h_{\mu\nu}T^{\mu\nu}$, see Eq (4.1) of \cite{GHL06},  where  $T^{\mu\nu}$  has the well-known form proportional to $u^\mu u^\nu$, see Eq. (4.3) of \cite{GHL06}.

For Case C under study here, the interaction action is given by Eq. (\ref{15}): the system is the geodesic deviation vector $ z^i$, the graviton field has two polarizations each represented by a massless minimally-coupled scalar field.  The second equality in Eq. (\ref{15}) obtains after an integration by parts on the $\ddot h$ thus, passing one time derivative to each $ z$, thus ending up with the same form as the $T^{\mu\nu}$ for the mass, denoted by a field  $X$ in Eq. (\ref{15}), and an interaction action of the form $h_{\mu\nu} T^{\mu\nu}$ as in Case B. 

The stress energy tensor of the matter field is obtained by taking the variational derivative of the effective action with respect to the metric function while the equation of motion for the system variables $h_{\mu\nu}$  is obtained by taking the variational derivative with respect to $T^{\mu\nu}$. Note that to get the dynamics of the expectation values one needs to use the in-in, CTP or Schwinger-Keldysh effective action. This was done in \cite{CalHu87,CamVer96} for the weak anistropy and  inhomogeneities cases. 

%it is the stress energy tensor of the gravitons; for B) C), that of the quantum field. [ For B) the expectation value  is the primary source of radiation reaction because usually one doesn't (or the time hasn't come yet for one to) care so much about their fluctuations.  ] 

\subsubsection{The fluctuations of the environmental variables appearing as noise, acting effectively as classical stochastic source}

The reason why, and the way how, quantum fluctuations can be viewed as noise, and be treated as a  classical stochastic force  is made clear by Feynman and Vernon  (FV) for Gaussian systems. Note that even though a classical stochastic (CS) interpretation is possible, one should appreciate that the FV treatment is fully quantum field-(QF) theoretical. Fundamental distinctions remain  between the innate QF features and an effective CS treatment. Their differences will become more overtly discernible for quantum information issues involving quantum coherence and entanglement   (See, e.g., \cite{HHCosEnt,AHChannel} for a more in-depth explanation.)  

For A) the effects of quantum noise from  fluctuations of matter field on the background spacetime dynamics are contained in a stochastic equation of motion in the form of a Langevin equation derivable from a stochastic effective action. This equation is called the Einstein-Langevin equation \cite{CalHu94,HuMat95,HuSin95,LomMaz,CamVer96,MarVer}, with sources arising from both  the expectation values of the stress energy tensor of the matter quantum field and its fluctuations,  measured by the stress energy two point function (the vacuum expectation value of the stress energy bitensor \cite{PH11}), known as the noise kernel.  While the expectation values of the stress energy tensor of the matter quantum field can be obtained by taking the first functional derivative of the CTP effective action  with respect to the background metric \cite{CalHu87,CamVer94}, the noise kernel can be obtained from its second functional derivative  of the CTP effective action, or the Euclidean action in an Euclidean formulation \cite{PH97,ChoHu11}.   

In case B) the effects of quantum noise from  fluctuations of quantum field on the trajectory of a charge or mass  are contained in a stochastic equation of motion in the form of a Langevin equation derivable from a stochastic effective action.
For charge, this is called the stochastic Abraham-Lorenz-Dirac (ALD) equation \cite{JH1,QRad}; for mass, the stochastic Mino-Sasaki-Tanaka-Quinn-Wald (MST-QW)  equation.

\subsubsection{Issues}

\paragraph{Fluctuation-Dissipation Relation (FDR)}
The linkage between the system and the environment manifests in the open system framework as dissipation in the system caused by the fluctuations in the environment.  For linear systems the dissipation and fluctuation kernels of a quantum field are embodied in the retarded Green function and the Hadamard function respectively,  corresponding to the  commutator and anti-commutator of the quantum field \cite{RHK}. They are related by a FDR relation. A FDR exists in Case C treated here, as we will show below. 

FDR is a condition of self-consistency and it is predicated on a stationarity condition.  This FDR  in the quantum field can induce a set of FDRs in the system. This has been shown for masses or atoms moving in a quantum field \cite{RHA,CPR} in models based on quantum Brownian oscillators in a fixed bath, like an unsqueezed quantum field. For a squeezed quantum field,   a FDR can still exist in the field \cite{HHFDRSq} but there is no FDR for an atom (or Unruh-DeWitt detector) interacting with the squeezed quantum field,  because it does not reach a stationarity condition at late times.    However, from an energy conservation consideration, one can show that the power generated in the dissipative dynamics of the system is indeed equal to the rate of energy produced associated with the particles created \cite{CalHu87}, and the fluctuations in the particle number is what enters in the noise kernel \cite{HuSin95,CamVer96}.   

\paragraph{Gravitons versus gravitational waves}
For gravitons with wavelengths much shorter than the geodesic separation, under the Brill-Hartle-Isaacson average, they act collectively as a  radiation fluid with equation of state where the pressure $p = \frac{1}{3}\rho$ the energy density, exactly like the corresponding case  for  photons. The geodesic deviation of the two masses will not be affected by the average value of these very short wavelength modes.  Notice that gravitons are the quantized short wavelength modes of weak perturbations. (The very long  wavelength modes are more wavelike, albeit with quantum properties. We would not exclude the possibility of the detection in the future of these `infrared gravitons' in a new field known as {\it gravitonics},  similar to   the recently observed infrared single photon phenomena \cite{IRphoton} in photonics.)  The   averaged value of the stress energy tensor  carried by these gravitational waves are evaluated in the same way as one would for a LIGO detector.  Beneath this is the interesting question of when, or how large,  a bunch of gravitons will show wavelike behavior. Leaving aside the decoherence and quantum to classical transition, this question would exactly be like that posed to photons in relation to EM waves, whose answer we know well. 

\paragraph{Gravitational Decoherence}
Gravitational decoherence is a class of problems of fundamental significance and of increasing current interest \cite{AHGravDec,Blencowe,LagAna,GravDecRev,GravDecTO},
where thermal gravitons (quantized weak gravitational perturbations) or gravitational field act as an environment in  decohering some quantum systems of interest.  E.g., the results for the thermal graviton polarization tensor   \cite{CamHu} obtained from the CTP effective action  and the Langevin equations derived by the influence functional method have been fruitfully applied to the derivation of a master equation of gravitational decoherence by Blencowe \cite{Blencowe}. (The same equation is obtained via canonical quantization in Ref  \cite{AHGravDec}). 
Decoherence by gravitons is also treated recently by Kanno et al \cite{Kanno}.

%With these precedent examples it is easy to see the flow of logic and the technical structure of our present problem.  We now proceed with the exposition of our calculations. 

\section{Stochastic influence action and Langevin equations}

We adopt  the Feynman-Vernon influence functional formalism \cite{FeyVer63} to investigate the influence of  gravitons on a skeletal interferometer represented by two masses, as done in \cite{PWZ21a,PWZ21b}. We begin with the actions for the gravitons and the masses, as well as the interaction term between them, and arrive at the stochastic effective action. From this we derive the Langevin equation of motion for the geodesic separation between the masses. Concentrating on the noise effect, the separation as well as its fluctuations are then solved perturbatively.

\subsection{Graviton action}

To obtain the action for the gravitons, we start with the Einstein action
\begin{eqnarray}
S_{g}=\frac{1}{\kappa^{2}}\int d^{4}x \sqrt{-g}\,R
\end{eqnarray}
where $\kappa^{2}=16\pi G$ and $G$ is the Newton's constant. Consider gravitational perturbations around the Minkowski background
\begin{eqnarray}
g_{\mu\nu}=\eta_{\mu\nu}+\kappa h_{\mu\nu}
\end{eqnarray}
Neglecting graviton self-interactions, we expand the Einstein action up to $O(h^{2})$. The expression can be simplified by imposing the transverse-traceless gauge,
\begin{eqnarray}
h_{0\mu}=0\ \ \ ;\ \ \ {h_{\mu}}^{\mu}=0\ \ \ ;\ \ \ \partial^{\mu}h_{\mu\nu}=0\label{TTgauge}
\end{eqnarray}
leading to  the graviton action
\begin{eqnarray}
S_{grav}= -\frac{1}{4}\int d^{4}x\,(\partial_{\alpha}h_{\mu\nu})(\partial^{\alpha}h^{\mu\nu})
\end{eqnarray}
Expressing the graviton field in terms of its two physical degrees of freedom corresponding to the two polarization components,
\begin{eqnarray}
h_{\mu\nu}(x)=\int d^{3}k\sum_{s}\epsilon_{\mu\nu}^{(s)}(\vec{k})\,h^{(s)}(\vec{k},t)\,e^{i\vec{k}\cdot\vec{x}}\label{gracom}
\end{eqnarray}
Since $h_{\mu\nu}(x)$ is real, we have 
\begin{eqnarray}
{\epsilon^{(s)}_{\mu\nu}}^{*}(\vec{k})=\epsilon^{(s)}_{\mu\nu}(-\vec{k})\ \ \ ;\ \ \ 
{h^{(s)}}^{*}(\vec{k},t)=h^{(s)}(-\vec{k},t)
\end{eqnarray}
Then the graviton action can be rewritten as
\begin{eqnarray}
S_{grav}=-\frac{1}{2}\int\,d^{4}x\,\sum_{s}\partial_{\alpha}h^{(s)}(x)\partial^{\alpha}h^{(s)}(x)\label{gaction}
\end{eqnarray}
where
\begin{eqnarray}
h^{(s)}(x)=\int\,d^{3}k\,h^{(s)}(\vec{k},t)\,e^{i\vec{k}\cdot\vec{x}}
\end{eqnarray}
In arriving at the result in Eq.~(\ref{gaction}), we have used the relation $\epsilon_{ij}^{(s)}=\sqrt{2}\epsilon_{i}^{(s)}\epsilon_{j}^{(s)}$ for circular polarizations, and the orthogonality of the polarization vectors $\hat{\epsilon}^{(s)*}\cdot\hat{\epsilon}^{(s')}=\delta_{ss'}$. This is the well-known result in which the graviton field can be treated as two massless minimally-coupled scalar fields in Robertson-Walker spacetimes with the Minkowski spacetime as a special case \cite{FP77}.

\subsection{Matter action}

Next, we consider the matter action in which there are two test masses. It is convenient to set up a Fermi normal coordinate system $(t,\vec{ z})$ \cite{MM63} along the geodesic of the first mass with the metric
\begin{eqnarray}
g_{00}(t,\vec{ z})&=&-1-R_{i0j0}(t,0) z^{i} z^{j}+\cdots\\
g_{0i}(t,\vec{ z})&=&-\frac{2}{3}R_{0ijk}(t,0) z^{j} z^{k}+\cdots\\
g_{ij}(t,\vec{ z})&=&\delta_{ij}-\frac{1}{3}R_{ikjl}(t,0) z^{k} z^{l}+\cdots
\end{eqnarray}
where $\vec{ z}$ is the geodesic deviation vector. Note that the transverse-traceless (TT) gauge used in Eq.~(\ref{TTgauge}) and the Fermi normal coordinate system here are compatible at the level of approximation used in our analysis, although there are subtle differences between the use of gauge and coordinates \footnote{Even though `gauge’ and `coordinates’ are often thought of as interchangeable,  there are subtle differences in the physical contexts and meanings used (at least in the stricter general relativity community). E.g., the TT gauge has the important physical significance of carrying the true dynamical degrees of freedom of gravitation (like the Lorentz gauge in electromagnetism), but not the Newton gauge (or Coulomb gauge in electromagnetism) which is `pure gauge’.  When it comes  to coordinate choices, they are usually adopted for some particular purpose or usage, e.g., the Riemann normal coordinate (RNC) is a quasi-local coordinate patch going beyond the local Lorentz frame, where the Riemann curvature tensor begins to appear. Fermi normal coordinate is defined as a RNC along a worldline.}. In terms of the gravitational perturbation $\kappa h_{\mu\nu}$, the Riemann tensor component
\begin{eqnarray}
R_{i0j0}=-\frac{\kappa}{2}\,\ddot{h}_{ij}\label{riecom}
\end{eqnarray}
The action of the second mass, with mass $m$, is just
\begin{eqnarray}
S_{m}=-m\int\,\sqrt{-ds^{2}}
\end{eqnarray}
Take the coordinates of the second mass be $(t,\vec{ z})$. Expanding the matter action $S_{m}$ to second order in $\vec{ z}$ and neglecting a constant term as well as boundary terms,
\begin{eqnarray}
S_{m}=\int\,dt\,\left[\frac{m}{2}\,\delta_{ij}\,\dot{ z}^{i}\dot{ z}^{j}+\frac{m\kappa}{4}\,\ddot{h}_{ij}\, z^{i} z^{j}\right]+\cdots\label{maction}
\end{eqnarray}
where we have expressed the Riemann curvature component in terms of the gravitational perturbation.

The first term in Eq.~(\ref{maction}) can be viewed as a kinetic term, while the second term as the interaction between the mass and the graviton. Since the graviton field can be decomposed in terms of its polarization components as shown in Eq.~(\ref{gracom}), we substitute this into the interaction term.
\begin{eqnarray} \label{15}
\int\,dt\,\frac{m\kappa}{4}\,\ddot{h}_{ij}\, z^{i} z^{j}=\alpha\int\,d^{4}x\,\sum_{s}h^{(s)}(x)X^{(s)}(x)
\end{eqnarray}
where the constant $\alpha=m\kappa/2\sqrt{2}(2\pi)^3$ and the source term is defined as
\begin{eqnarray}
X^{(s)}(x)=\int\,d^{3}k\,e^{i\vec{k}\cdot\vec{x}}\frac{d^{2}}{dt^{2}}\left({\epsilon^{(s)}_{i}}^{*}(\vec{k}) z^{i}(t)\right)^{2}\label{defX}
\end{eqnarray}
We see that the two polarization components also decouple in the interaction term. Hence, we can consider the effect of each graviton polarization separately.

\subsection{Stochastic effective action}

Let us recapitulate. Adding the graviton and the matter actions together, we have the total action
\begin{eqnarray}
S_{total}&=&S_{g}+S_{m}\nonumber\\
&=&\frac{m}{2}\int\,dt\,\delta_{ij}\,\dot{ z}^{i}\dot{ z}^{j}+\sum_{s}\int\,d^{4}x\,\left[-\frac{1}{2}\,\partial_{\alpha}h^{(s)}(x)\partial^{\alpha}h^{(s)}(x)+\alpha h^{(s)}(x)X^{(s)}(x)\right]+\cdots,\nonumber\\
\end{eqnarray}
where the integration over $t$ is from some initial time, which we shall take to be $t=0$, to $\infty$. As we have stated above, the two graviton polarizations decouple and one can consider them one at a time. To study the influence of the gravitons on the mass $m$, one needs to calculate the various in-in expectation values of the graviton field. This can be achieved by the closed-time path integral formalism. In the present case, if we define $J_{\pm}(x)=\alpha X_{\pm}(x)$, the integration over the gravitons, for one polarization $h(x)$, can be written as \cite{CalHu87,CalHu88,CalHu08}
\begin{eqnarray}
e^{iS_{IF}}&=&\int_{CTP}Dh_{+}\,Dh_{-}\,e^{i(S_{g}[h_{+}]-S_{g}[h_{-}]+\int J_{+}h_{+}-\int J_{-}h_{-})}\nonumber\\
&=&e^{-\frac{i}{2}\int (J_{+}G_{++}J_{+}-J_{+}G_{+-}J_{-}-J_{-}G_{-+}J_{+}+J_{-}G_{--}J_{-})}\label{IFaction1}
\end{eqnarray}
where we have defined the influence action $S_{IF}$. The Schwinger-Keldysh Green functions in this expression are
\begin{eqnarray}
G_{++}(x,x')&=&-i\langle Th_{+}(x)h_{+}(x')\rangle\\
G_{+-}(x,x')&=&-i\langle h_{-}(x')h_{+}(x)\rangle\\
G_{-+}(x,x')&=&-i\langle h_{-}(x)h_{+}(x')\rangle=G_{+-}(x',x)\\
G_{--}(x,x')&=&-i\langle \bar{T}h_{-}(x)h_{-}(x')\rangle=-G_{++}^{*}(x,x')
\end{eqnarray}
where $T$ and $\bar{T}$ represent the time ordered and anti-time ordered operations, respectively. In the evaluation of these Green functions, one could take the expectation values in the vacuum state as well as in the thermal, coherent, and squeezed states according to the physical situation involved. We shall discuss more on this in the following sections. 

The influence action $S_{IF}$ can be further simplified by defining
\begin{eqnarray}
\Sigma^{ij}(t)&=&\frac{1}{2}\left[ z_{+}^{i}(t) z_{+}^{j}(t)+ z_{-}^{i}(t) z_{-}^{j}(t)\right]\\\Delta^{ij}(t)&=& z_{+}^{i}(t) z_{+}^{j}(t)- z_{-}^{i}(t) z_{-}^{j}(t)
\end{eqnarray}
Then, recalling the definition of $X(x)$, the influcence action can be written as
\begin{eqnarray}
S_{IF}&=&\int dt\,dt'\,\Delta^{ij}(t)D_{ijkl}(t,t')\Sigma^{kl}(t')+\frac{i}{2}\int dt\, dt'\,\Delta^{ij}(t)N_{ijkl}(t,t')\Delta^{kl}(t')\label{IFaction2}
\end{eqnarray}
Here,
\begin{eqnarray}
D_{ijkl}(t,t')=\alpha^{2}\frac{d^{2}}{dt^{2}}\frac{d^{2}}{dt'^{2}}\int d^{3}k\,d^{3}k'\int d^{3}x\,d^{3}x'\,e^{-i\vec{k}\cdot\vec{x}}e^{-i\vec{k}'\cdot\vec{x}'}\sum_{s}\epsilon_{ij}^{(s)}(\vec{k})\epsilon_{kl}^{(s)}(\vec{k}')G_{ret}(x,x')\nonumber\\
\end{eqnarray}
where 
\begin{eqnarray}
G_{ret}(x,x')=i\theta(t-t')\langle[h(x),h(x')]\rangle
\end{eqnarray}
is the retarded Green function. $D_{ijkl}(t,t')$ is usually called the dissipation kernel. Also,
\begin{eqnarray}
N_{ijkl}(t,t')=\frac{\alpha^{2}}{2}\frac{d^{2}}{dt^{2}}\frac{d^{2}}{dt'^{2}}\int d^{3}k\,d^{3}k'\int d^{3}x\,d^{3}x'\,e^{-i\vec{k}\cdot\vec{x}}e^{-i\vec{k}'\cdot\vec{x}'}\sum_{s}\epsilon_{ij}^{(s)}(\vec{k})\epsilon_{kl}^{(s)}(\vec{k}')G^{(1)}(x,x')\nonumber\\
\label{noise}
\end{eqnarray}
where 
\begin{eqnarray}
G^{(1)}(x,x')=\langle\{h(x),h(x')\}\rangle
\end{eqnarray}
is the Hadamard function. $N_{ijkl}(t,t')$ is called the noise kernel. The kernels $D_{ijkl}(t,t')$ and $N_{ijkl}(t,t')$ are related by the fluctuation-dissipation relation. Note that we have included the effect of both polarizations in the final form of the influence action in Eq.~(\ref{IFaction2}).

The term quadratic in $\Delta^{ij}(t)$ in $S_{IF}$ can be formulated in terms of a stochastic tensor $\xi_{ij}(t)$ using the Feynman-Vernon Gaussian functional identity.
\begin{eqnarray}
e^{-\frac{1}{2}\int\,\Delta^{ij}N_{ijkl}\Delta^{kl}}
&=&{\cal N}\int\ D\xi\ e^{-\frac{1}{2}\int\,(\xi_{ij}+i\Delta^{mn}N_{ijmn})(N^{-1})^{ijkl}(\xi_{kl}+iN_{klpq}\Delta^{pq})}\ e^{-\frac{1}{2}\int\,\Delta^{ij}N_{ijkl}\Delta^{kl}}\nonumber\\
&=&{\cal N}\int\ D\xi\ e^{-\frac{1}{2}\int\,\xi_{ij}(N^{-1})^{ijkl}\xi_{kl}}\ e^{-i\int\,\xi_{ij}\Delta^{ij}}
\end{eqnarray}
where ${\cal N}$ is a normalization constant. The influence action then becomes
\begin{eqnarray}
e^{iS_{IF}}=\int\,D\xi\ P[\xi]\ e^{i\int\,\Delta^{ij}D_{ijkl}\Sigma^{kl}-i\int\,\xi_{ij}\Delta^{ij}}
\end{eqnarray}
where the gaussian probability density
\begin{eqnarray}
P[\xi]={\cal N}\, e^{-\frac{1}{2}\int\,\xi_{ij}(N^{-1})^{ijkl}\xi_{kl}}
\end{eqnarray}
Stochastic averages are performed using this probability density. For example, the two point correlation function
\begin{eqnarray}
\langle\xi_{ij}(t)\xi_{kl}(t')\rangle_{s}
&=&\int\,D\xi\ P[\xi]\,\xi_{ij}(t)\xi_{kl}(t')\nonumber\\
&=&N_{ijkl}(t,t')
\end{eqnarray}
That is the reason why $N_{ijkl}(t,t')$ is called the noise kernel.

Together with the matter actions, one can construct the stochastic effective action
\begin{eqnarray}
S_{SEA}&=&S_{m}[ z_{+}]-S_{m}[ z_{-}]+S_{IF}\nonumber\\
&=&\frac{m}{2}\int dt\,\delta_{ij}\dot{ z}_{+}^{i}(t)\dot{ z}_{+}^{j}(t)-\frac{m}{2}\int dt\,\delta_{ij}\dot{ z}_{-}^{i}(t)\dot{ z}_{-}^{j}(t)\nonumber\\
&&\hskip 10pt +\int dt\,dt'\,\Delta^{ij}(t)D_{ijkl}(t,t')\Sigma^{kl}(t')-\int dt\,\xi_{ij}(t)\Delta^{ij}(t)
\end{eqnarray}
From this effective action, one can derive the corresponding equation of motion for $ z(t)$ under the effect of the stochastic tensor force $\xi_{ij}(t)$ in the form of a Langevin equation.

\subsection{Langevin equation}

The equation of motion for $ z^{i}(t)$ can now be obtained by taking the variation on the stochastic effective action
\begin{eqnarray}
&&\left.\frac{\delta S_{SEA}}{\delta z^{i}_{+}}\right|_{ z_{+}= z_{-}= z}=0\nonumber\\
&\Rightarrow& m\ddot{ z}^{i}(t)+2\delta^{im}\int dt'\,D_{mnkl}(t,t')\, z^{n}(t) z^{k}(t') z^{l}(t')-2\delta^{ik}\xi_{kl}(t)\, z^{l}(t)=0\label{langevin1}
\end{eqnarray}
For the investigation of the fluctuations or noise of gravitons one does not need to carry a classical gravitational wave, which is what we assumed. Hence, we do not expect a linear term to appear in Eq.~(\ref{langevin1}). It is like the use of a background field expansion in quantum field theory but assuming that the classical background field to be absent.  Of course there is no harm in carrying it along, as PWZ did.  

Note that the second term in Eq.~(\ref{langevin1}) involving the dissipation kernel is history dependent. This is a nonlinear integral differential equation which is very difficult to solve analytically. 
In the following we shall use a perturbative approach. Without the graviton effects, we have the homogeneous equation,
\begin{eqnarray}
m\ddot{ z}_{0}^{i}(t)=0
\end{eqnarray}
where $ z_{0}^{i}(t)$ corresponds to the geodesic motion in the background spacetime. In this perturbative approach, the next order effect comes from the noise term which is linear in $ z^{i}(t)$. Suppose we take $ z^{i}= z_{0}^{i}+\delta z^{i}$. 
Then,
\begin{eqnarray}
m\,\ddot{\delta z}^{i}(t)= 2\delta^{ik}\xi_{kl}(t)\, z_{0}^{l}(t)+\cdots\label{langevin2}
\end{eqnarray}
Imposing the initial conditions,
\begin{eqnarray}
\delta z^{i}(0)=\dot{ z}^{i}(0)=0,
\end{eqnarray}
the solution $\delta z(t)$ can be written as
\begin{eqnarray}
\delta z^{i}(t)=\frac{2}{m}
\int_{0}^{t}dt'\,(t-t')\delta^{ik}\xi_{kl}(t') z_{0}^{l}(t').
\end{eqnarray}
This represents the fluctuation of the mass $m$ due to the noise term coming from the graviton effect. The correlator of this fluctuation
\begin{eqnarray}
\langle\delta z^{i}(t)\delta z^{j}(t')\rangle_{s}
&=&\frac{4}{m^{2}}\delta^{ik}\delta^{jl}\int_{0}^{t}dt''\int_{0}^{t'}dt'''(t-t'')(t'-t''') z_{0}^{m}(t'') z_{0}^{n}(t''')N_{kmln}(t'',t'''),\nonumber\\
\label{sepflu}
\end{eqnarray}
which is directly related to the noise kernel. Moreover, the fluctuation of $\delta\xi(t)$ itself is given by the self correlation of $\delta\xi(t)$. Due to the importance of the noise kernel in the understanding of these correlations, we shall evaluate this kernel in the next section for various quantum state of the graviton including the Minkowski vacuum, the thermal, the coherent, and the squeezed states \cite{Caves}.

\section{Noise kernels of different quantum states}\label{noisekernel}

As we have stated above, the deviation $\delta z^{i}(t)$ from the geodesic motion due to the noise effect of the graviton is closely related to the noise kernel defined in Eq.~(\ref{noise}). In turn the noise kernel is expressed in terms of the Hadamard function $G^{(1)}(x,x')$ of the graviton. Depending on the quantum state the graviton is in, its corresponding Hadamard functions would have different properties. Hence, in this section we shall work out in some detail the noise kernels in the Minkowski vacuum, thermal, coherent, and squeezed states. 

\subsection{Minkowski vacuum}

For the Minkowski vacuum, the mode function can be written as
\begin{eqnarray}
u_{\vec{k}}(x)=\frac{1}{(2\pi)^{3/2}\sqrt{2k}}e^{-ikt}e^{i\vec{k}\cdot\vec{x}},\label{modfun}
\end{eqnarray}
where for massless particles the frequency of the $\vec{k}$ th normal mode  is $\omega = k \equiv |\vec k|$. And, in terms of the mode function, the Hadamard function can be expressed as
\begin{eqnarray}
G^{(1)}(x,x')&=&\int\,d^{3}k\left[u_{\vec{k}}(x)u_{\vec{k}}^{*}(x')+u_{\vec{k}}^{*}(x)u_{\vec{k}}(x')\right].
\end{eqnarray}
With this Hadamard function, the noise kernel in Eq.~(\ref{noise}) can be simplified to
\begin{eqnarray}
N_{ijkl}^{(0)}(t,t')&=&4\pi^{3}\alpha^{2}\frac{d^{2}}{dt^{2}}\frac{d^{2}}{dt'^{2}}\int\,d^{3}q\left(\frac{\cos[q(t-t')]}{q}\right)\sum_{s}\epsilon_{ij}^{(s)}(\vec{q}){\epsilon_{kl}^{(s)}}^{*}(\vec{q}).
\end{eqnarray}
The superscript $(0)$ denotes that the noise kernel is evaluated using the Minkowski vacuum state. The polarization sum gives the projection operator $P_{ijkl}(\vec{q})$ (see Eq.~(\ref{polsum}) and other related formulae listed in the Appendix). The angular integration over $\vec{q}$ can be performed as listed in Eq.~(\ref{intpijkl}). The noise kernel can then be expressed as an integration over $q=|\vec{q}|$.
\begin{eqnarray}
N_{ijkl}^{(0)}(t,t')=-\frac{32\pi^{4}\alpha^{2}}{15}[2\,\delta_{ij}\delta_{kl}-3(\delta_{ik}\delta_{jl}+\delta_{il}\delta_{jk})]\int_{0}^{\infty}dq\,q^{5}\cos[q(t-t')].
\end{eqnarray}
The integral over $q$ above is divergent and must be regularized. Here we shall use a momentum cutoff $\Lambda$ which will be related to some scale in the problem. 
In the present case, as we have commented on this issue in Sec.~II, this scale should be of the order of $1/z_{0}$, where $z_{0}$ is the initial geodesic separation between the two masses. We shall discuss more on this when we estimate the detectability of the noise effect of gravitons in the next section.

With the cutoff $\Lambda$,
\begin{eqnarray}
\int_{0}^{\Lambda}dq\,q^{5}\,\cos[q(t-t')]=\Lambda^{6}F[\Lambda(t-t')],
\end{eqnarray}
where 
\begin{eqnarray}
F(x)&=&\frac{1}{x^{6}}\int_{0}^{x}dy\,y^{5}\cos y\nonumber\\
&=&\frac{1}{x^{6}}\left[(5x^{4}-60x^{2}+120)\cos x+x(x^{4}-20x^{2}+120)\sin x-120\right],\label{defF}
\end{eqnarray}
which is the same function introduced in \cite{Kanno}. Hence, the noise kernel in the Minkowski vacuum state is 
\begin{eqnarray}
N_{ijkl}^{(0)}(t,t')=-\left(\frac{32\pi^{4}}{15}\right)\alpha^{2}\Lambda^{6}\,[2\,\delta_{ij}\delta_{kl}-3(\delta_{ik}\delta_{jl}+\delta_{il}\delta_{jk})]\,F[\Lambda(t-t')].\label{vacnoi}
\end{eqnarray}
For small $\Lambda(t-t')$, 
\begin{eqnarray}
N_{ijkl}^{(0)}(t,t')=-\left(\frac{16\pi^{4}}{45}\right)\alpha^{2}\Lambda^{6}\,[2\,\delta_{ij}\delta_{kl}-3(\delta_{ik}\delta_{jl}+\delta_{il}\delta_{jk})]\left[1-\frac{3}{8}\Lambda^{2}(t-t')^{2}+\cdots\right].
\end{eqnarray}
For large $\Lambda(t-t')$,
\begin{eqnarray}
N_{ijkl}^{(0)}(t,t')&=&-\left(\frac{32\pi^{4}}{15}\right)\alpha^{2}\Lambda^{6}\,[2\,\delta_{ij}\delta_{kl}-3(\delta_{ik}\delta_{jl}+\delta_{il}\delta_{jk})]\nonumber\\
&&\ \ \ \ \ \left\{\frac{\sin[\Lambda(t-t')]}{\Lambda(t-t')}+\frac{5\,\cos[\Lambda(t-t')]}{\Lambda^{2}(t-t')^{2}}+\cdots\right\}
\end{eqnarray}

\subsection{Thermal state}

In this subsection, we consider the finite temperature case with the thermal state characterized by the temperature $T$ or $\beta=1/T$. Here the Hadamard function is
\begin{eqnarray}
G^{(1)}_{\beta}(x,x')=\sum_{n=-\infty}^{\infty}\frac{1}{2\pi^{2}}\left[\frac{1}{|\vec{x}-\vec{x}'|^{2}-(t-t'+in\beta)^{2}}\right]
\end{eqnarray}
where $n=0,\pm1, \pm2, \dots$. The $n=0$ term is just the Minkowski vacuum part derived above. It is divergent and regularization has to be implemented as indicated there. Other than this term, one can identify the rest to be the thermal part of the noise kernel, 
\begin{eqnarray}
N_{ijkl}^{(\beta)}(t,t')&=&\left(\frac{\alpha^{2}}{4\pi^{2}}\right)\frac{d^{2}}{dt^{2}}\frac{d^{2}}{dt'^{2}}\int d^{3}k\,d^{3}k'\int d^{3}x\,d^{3}x'\,e^{-i\vec{k}\cdot\vec{x}}e^{-i\vec{k}'\cdot\vec{x}'}\nonumber\\
&&\ \ \ \ \ \sum_{s}\epsilon_{ij}^{(s)}(\vec{k})\epsilon_{kl}^{(s)}(\vec{k}')
\sum_{n=-\infty}^{\infty}\!\!{\vphantom{\sum}}'\left[\frac{1}{|\vec{x}-\vec{x}'|^{2}-(t-t'+in\beta)^{2}}\right]\label{thermalN}
\end{eqnarray}
where the prime denotes that the $n=0$ term has been left out.

As the Hadamard function depends only on $\vec{x}-\vec{x}'$, one can define a new variable $\vec{y}=\vec{x}-\vec{x}'$ with $\int\,d^{3}x\,d^{2}x'\rightarrow\int\,d^{3}y\,d^{3}x'$. After the integrations and the sum over polarizations one obtains
\begin{eqnarray}
N_{ijkl}^{(\beta)}(t,t')&=&-\frac{64\pi^{3}\alpha^{2}}{15}[2\,\delta_{ij}\delta_{kl}-3(\delta_{ik}\delta_{jl}+\delta_{il}\delta_{jk})]\nonumber\\
&&\ \ \frac{d^{2}}{dt^{2}}\frac{d^{2}}{dt'^{2}}\int_{0}^{\infty}dk\,k\int_{0}^{\infty}dy\,y\sin(ky){\sum_{n=-\infty}^{\infty}}\!{\vphantom{\sum}}'\left[\frac{1}{y^{2}-(t-t'+in\beta)^{2}}\right].
\end{eqnarray}
The integration over $y$ can be evaluated readily using, for example, the residue method. Then, summing over $n$, we have
\begin{eqnarray}
\int_{0}^{\infty}dy\,y\sin(ky){\sum_{n=-\infty}^{\infty}}\!{\vphantom{\sum}}'\left[\frac{1}{y^{2}-(t-t'+in\beta)^{2}}\right]=\frac{\pi\cos[k(t-t')]}{e^{k\beta}-1}.
\end{eqnarray}
Finally,
\begin{eqnarray}
N_{ijkl}^{(\beta)}(t,t')&=&-\frac{64\pi^{}\alpha^{2}}{15}[2\,\delta_{ij}\delta_{kl}-3(\delta_{ik}\delta_{jl}+\delta_{il}\delta_{jk})]\,\frac{d^{2}}{dt^{2}}\frac{d^{2}}{dt'^{2}}
\int_{0}^{\infty}dk\,k\,\frac{\cos[k(t-t')]}{e^{k\beta}-1}\nonumber\\
&=&-256\pi^{10}\alpha^{2}\,[2\,\delta_{ij}\delta_{kl}-3(\delta_{ik}\delta_{jl}+\delta_{il}\delta_{jk})]\nonumber\\
&&\ \ \ \ \ \left(\frac{1}{\beta^{6}x^{6}}\right)\left[1-\frac{x^{6}}{15\sinh^{6}x}\left(2+11\cosh^{2}x+2\cosh^{4}x\right)\right],\label{theNK}
\end{eqnarray}
where $x=\pi(t-t')/\beta$.

At low temperature, that is, $x\ll 1$ or $\beta\gg \pi(t-t')$,
\begin{eqnarray}
N_{ijkl}^{(\beta)}(t,t')&=&-\frac{512\pi^{10}\alpha^{2}}{945}\,[2\,\delta_{ij}\delta_{kl}-3(\delta_{ik}\delta_{jl}+\delta_{il}\delta_{jk})]\left(\frac{1}{\beta^{6}}\right)\left[1-\frac{21}{10}x^{2}+\frac{14}{11}x^{4}+\cdots\right].\label{lowNK}\nonumber\\
\end{eqnarray}
At high temperature, $x\gg 1$ or $\beta\ll\pi(t-t')$,
\begin{eqnarray}
N_{ijkl}^{(\beta)}(t,t')&=&-256\pi^{10}\alpha^{2}\,[2\,\delta_{ij}\delta_{kl}-3(\delta_{ik}\delta_{jl}+\delta_{il}\delta_{jk})]\nonumber\\
&&\ \ \ \ \ \left(\frac{1}{\beta^{6}}\right)\left[\frac{1}{x^{6}}-\frac{8}{15}e^{-2x}-\frac{176}{15}e^{-4x}-\frac{128}{15}e^{-6x}+\cdots\right].\label{highNK}
\end{eqnarray}

\subsection{Coherent state}

The coherent state in field theory is defined by
\begin{eqnarray}
|\tilde{\alpha}_{\vec{k}}\rangle=D(\tilde{\alpha}_{\vec{k}})|0\rangle,
\end{eqnarray}
where the displacement operator (carrying an $\tilde{\alpha}$ argument to distinguish from the dissipation kernel appeared earlier)
\begin{eqnarray}
D(\tilde{\alpha}_{\vec{k}})=\prod_{\vec{k}}e^{\tilde{\alpha}_{\vec{k}}\hat{a}_{\vec{k}}^{\dagger}-\tilde{\alpha}_{\vec{k}}^{*}\hat{a}_{\vec{k}}}.
\end{eqnarray}
Usually, the parameter $\tilde{\alpha}$ is taken to be independent of $\vec{k}$. Then,
\begin{eqnarray}
D(\tilde{\alpha})=\prod_{\vec{k}}e^{\tilde{\alpha}\hat{a}_{\vec{k}}^{\dagger}-\tilde{\alpha}^{*}\hat{a}_{\vec{k}}}.
\end{eqnarray}
The corresponding Hadamard function in this state becomes
\begin{eqnarray}
G_{\tilde{\alpha}}^{(1)}(x,x')&=&\langle\tilde{\alpha}|\{\hat{h}(x),\hat{h}(x')\}|\tilde{\alpha}\rangle\nonumber\\
&=&\int\,d^{3}k\left[u_{\vec{k}}(x)u_{\vec{k}}^{*}(x')+u_{\vec{k}}^{*}(x)u_{\vec{k}}(x')\right]\nonumber\\
&&\ \ +2\int\,d^{3}k\,d^{3}k'\left[\tilde{\alpha}^{2}u_{\vec{k}}(x)u_{\vec{k}'}(x')+\tilde{\alpha}^{*2}u_{\vec{k}}^{*}(x)u_{\vec{k}'}^{*}(x')\right]\nonumber\\
&&\ \ +2\int\,d^{3}k\,d^{3}k'\,|\tilde{\alpha}|^{2}\left[u_{\vec{k}}(x)u_{\vec{k}'}^{*}(x')+u_{\vec{k}}^{*}(x)u_{\vec{k}'}(x')\right].
\end{eqnarray}
The first term above is just the Minkowski vacuum part. Hence, using the rest of the Hadamard function, one can define the coherent state part of the noise kernel as
\begin{eqnarray}
N_{ijkl}^{(\tilde{\alpha})}(t,t')
&=&\alpha^{2}\frac{d^{2}}{dt^{2}}\frac{d^{2}}{dt'^{2}}\int d^{3}k\,d^{3}k'\int d^{3}x\,d^{3}x'\,e^{-i\vec{k}\cdot\vec{x}}e^{-i\vec{k}'\cdot\vec{x}'}\sum_{s}\epsilon_{ij}^{(s)}(\vec{k})\epsilon_{kl}^{(s)}(\vec{k}')\nonumber\\
&&\ \ \int\,d^{3}q\,d^{3}q'\left[\tilde{\alpha}^{2}u_{\vec{q}}(x)u_{\vec{q}'}(x')+|\tilde{\alpha}|^{2}u_{\vec{q}}(x)u_{\vec{q}'}^{*}(x')+{\rm cc}\right],
\end{eqnarray}
where cc means complex conjugation.

The integration of the polarization tensor over solid angles is evaluated in the Appendix (Eq.~(\ref{defA})) expressed in terms of the tensor $A_{ij}$. Using this result and the explicit form of the mode function $u_{\vec{q}}(x)$ in Eq.~(\ref{modfun}), the coherent state part of the noise kernel can be evaluated to 
\begin{eqnarray}
N_{ijkl}^{(\tilde{\alpha})}(t,t')
&=&\left(\frac{64\pi^{5}}{9}\right)\alpha^{2}A_{ij}A_{kl}\frac{d^{2}}{dt^{2}}\frac{d^{2}}{dt'^{2}}\left[\tilde{\alpha}\int\,dk\,k^{3/2}e^{-ikt}+{\rm cc}\right]\left[\tilde{\alpha}\int\,dk'\,k'^{3/2}e^{-ik't'}+{\rm cc}\right].\nonumber\\
\end{eqnarray}
Now, if we take the coherent state parameter $\tilde{\alpha}$ to be real, the expression for $N_{ijkl}^{(\tilde{\alpha})}$ can be further simplified to
\begin{eqnarray}
N_{ijkl}^{(\tilde{\alpha})}=\left(\frac{256\pi^{5}}{9}\right)\alpha^{2}\tilde{\alpha}^{2}A_{ij}A_{kl}\int_{0}^{\infty}\,dk\,k^{7/2}\cos(kt)\int_{0}^{\infty}\,dk'\,k'^{7/2}\cos(k't').
\end{eqnarray}
The integrals over $k$ and $k'$ are divergent. As in the vacuum case, we shall put in a momentum cutoff $\Lambda$.
\begin{eqnarray}
\int_{0}^{\Lambda}dk\,k^{7/2}\cos(kt)=t^{-9/2}G(\Lambda t),
\end{eqnarray}
where the function
\begin{eqnarray}
G(x)=\frac{7}{8}\sqrt{x}(4x^{2}-15)\cos x+\frac{1}{4}x^{3/2}(4x^{2}-35)\sin x+\frac{105\sqrt{2\pi}}{16}\,C\left(\sqrt{\frac{2x}{\pi}}\right)\label{defG}
\end{eqnarray}
and $C(x)$ is the Fresnel integral
\begin{eqnarray}
C(x)=\int_{0}^{x}dt\cos\left(\frac{\pi t^{2}}{2}\right)
\end{eqnarray}
Finally, the coherent state part of the noise kernel comes down to
\begin{eqnarray}
N_{ijkl}^{(\tilde{\alpha})}=\left(\frac{256\pi^{5}}{9}\right)\,\alpha^{2}\tilde{\alpha}^{2}A_{ij}A_{kl}(tt')^{-9/2}G(\Lambda t)G(\Lambda t').\label{coherentNK}
\end{eqnarray}

\subsection{Squeezed state}

Similar to the coherent state, the squeezed state \cite{Caves} can be defined by
\begin{eqnarray}
|\zeta\rangle=S( \zeta)|0\rangle,
\end{eqnarray}
where the squeeze operator
\begin{eqnarray}
S( \zeta)=\prod_{\vec{k}}e^{\frac{1}{2} \zeta_{\vec{k}}^{*}\hat{a}_{\vec{k}}^{2}-\frac{1}{2} \zeta_{\vec{k}}\hat{a}_{\vec{k}}^{\dagger 2}},
\end{eqnarray}
and the squeeze parameter $ \zeta$ is taken to be independent of $\vec{k}$.

Taking the squeeze parameter $\zeta$ to be real, the corresponding Hadamard function is then given by
\begin{eqnarray}
G_{ \zeta }^{(1)}(x,x')
&=&\langle \zeta |\{\hat{h}(x),\hat{h}(x')\}| \zeta\rangle\nonumber\\
&=&(\cosh2 \zeta )\int\,d^{3}k\left[u_{\vec{k}}(x)u_{\vec{k}}^{*}(x')+u_{\vec{k}}^{*}(x)u_{\vec{k}}(x')\right]\nonumber\\
&&\ \ -(\sinh2 \zeta )\int\,d^{3}k\left[u_{\vec{k}}(x)u_{\vec{k}}(x')+u_{\vec{k}}^{*}(x)u_{\vec{k}}^{*}(x')\right].
\end{eqnarray}
The first term above is just $\cosh2 \zeta$ times the Minkowski vacuum Hadamard function. Therefore, the noise kernel in the squeezed state is given by
\begin{eqnarray}
N_{ijkl}^{( \zeta)}(t,t')
&=&(\cosh2 \zeta )N_{ijkl}^{(0)}(t,t')\nonumber\\
&&\ \ -(\sinh2 \zeta )\left(\frac{\alpha^{2}}{2}\right)\frac{d^{2}}{dt^{2}}\frac{d^{2}}{dt'^{2}}\int d^{3}k\,d^{3}k'\int d^{3}x\,d^{3}x'\,e^{-i\vec{k}\cdot\vec{x}}e^{-i\vec{k}'\cdot\vec{x}'}\nonumber\\
&&\ \ \ \ \ \ \ \sum_{s}\epsilon_{ij}^{(s)}(\vec{k})\epsilon_{kl}^{(s)}(\vec{k}')\int\,d^{3}q\left[u_{\vec{q}}(x)u_{\vec{q}}(x')+u_{\vec{q}}^{*}(x)u_{\vec{q}}^{*}(x')\right]
\end{eqnarray}
Let us concentrate on the second term. Again, using the formulae in the Appendix, especially Eq.~(\ref{intsum}), one can integrate the sum of the polarization tensors over the solid angles. Subsequently, the various integrations can be performed to obtain the following expression for this second term,
\begin{eqnarray}
-(\sinh2 \zeta )\left(\frac{16\pi^{4}}{15}\right)\alpha^{2}B_{ijkl}\int_{0}^{\infty}dq\,q^{5}\cos[q(t+t')].
\end{eqnarray}
Again, the integration over $q$ is divergent and a momentum cutoff $\Lambda$ will be implemented. With the function $F(x)$ defined in Eq.~(\ref{defF}), the $q$-integral is 
\begin{eqnarray}
\int_{0}^{\Lambda}dq\,q^{5}\cos[q(t+t')]=\Lambda^{6}F[\Lambda(t+t')].
\end{eqnarray}
With these considerations, the squeezed state noise kernel can finally be written as
\begin{eqnarray}
N_{ijkl}^{( \zeta )}(t,t')
&=&(\cosh2 \zeta )N_{ijkl}^{(0)}(t,t')\nonumber\\
&&\ \ -(\sinh2 \zeta )B_{ijkl}\left(\frac{16\pi^{4}}{15}\right)\alpha^{2}\Lambda^{6}F[\Lambda(t+t')].\label{squeezedNK}
\end{eqnarray}

\section{Graviton-induced geodesic separation and fluctuations}

With the noise kernels in various quantum states in the last section, we are ready to analyze the correlation function of the separation $\delta z^{i}$ as well as its fluctuations due to the influence of the gravitons in the form of the stochastic tensor force $\xi_{ij}$. These fluctuations are given by the expression in Eq.~(\ref{sepflu}). From these results, we can estimate the detectability of these fluctuations for various quantum states of the gravitons.

\subsection{Minkowski vacuum}

Using the noise kernel in the Minkowski vacuum in Eq.~(\ref{vacnoi}) and also Eq.~(\ref{sepflu}), the separation correlation can be written as
\begin{eqnarray}
\langle\delta z^{i}(t)\delta z^{j}(t')\rangle^{(0)}
&=&\left(\frac{128\pi^{4}}{15m^{2}}\right)\alpha^{2}\Lambda^{6}\left(3\,\delta_{ij}\delta_{kl}+\delta_{ik}\delta_{jl}\right) z_{0}^{k} z_{0}^{l}\nonumber\\
&&\ \ \ \ \ \int_{0}^{t}dt''\int_{0}^{t'}dt'''\,(t-t'')(t'-t''')F[\Lambda(t''-t''')]\,
\end{eqnarray}
where we have assumed $ z_{0}^{i}$ to be a constant vector. Henceforth we shall leave out the subscript $s$ denoting the stochastic average for notational simplicity. Although the $t$-integrations can be done in closed form, the final result is a bit lengthy and not very illuminating. It is convenient to express the result in power of $(t-t')$,
\begin{eqnarray}
&&\int_{0}^{t}dt''\int_{0}^{t'}dt'''\,(t-t'')(t'-t''')F[\Lambda(t''-t''')]\nonumber\\
&=&\frac{1}{4\Lambda^{6}t^{2}}\left[\Lambda^{4}t^{4}+4\Lambda^{2}t^{2}(1+2\cos(\Lambda t))-24\Lambda t\sin(\Lambda t)+24(1-\cos(\Lambda t))\right]\nonumber\\
&&\ \ +\frac{(t-t')}{4\Lambda^{6}t^{3}}\left[-\Lambda^{4}t^{4}+4\Lambda^{3}t^{3}\sin(\Lambda t)+12\Lambda^{2}t^{2}\cos(\Lambda t)-24\Lambda t\sin(\Lambda t)+24(1-\cos(\Lambda t))\right]\nonumber\\&&\ \ \ \ \ \ +\cdots.
\end{eqnarray}
Therefore, we can see that the coincident limit $t'\rightarrow t$ is finite for the correlator $\langle\delta z^{i}(t)\delta z^{j}(t')\rangle$. That is, the fluctuation of the separation $\delta z^{i}$ in the Minkowski vacuum is 
\begin{eqnarray}
\langle\delta z^{i}(t)\delta z^{j}(t)\rangle^{(0)}
&=&\left(3\,\delta_{ij}\delta_{kl}+\delta_{ik}\delta_{jl}\right) z_{0}^{k} z_{0}^{l}\left(\frac{1}{240\pi^{2}t^{2}}\right)\nonumber\\
&&\ \ \left[\Lambda^{4}t^{4}+4\Lambda^{2}t^{2}(1+2\cos(\Lambda t))-24\Lambda t\sin(\Lambda t)+24(1-\cos(\Lambda t))\right].\nonumber\\
\label{vacflu}
\end{eqnarray}
where we have substituted $\alpha=m\kappa/2\sqrt{2}(2\pi)^{3}$ into the expression above.
If we set, without loss of generality, $ z_{0}^{i}=(0,0, z_{0})$, then
\begin{eqnarray}
\langle(\delta z^{3}(t))^{2}\rangle^{(0)}
&=&\left(\frac{\kappa^{2} z_{0}^{2}}{60\pi^{2}t^{2}}\right)
[\Lambda^{4}t^{4}+4\Lambda^{2}t^{2}(1+2\cos(\Lambda t))\nonumber\\
&&\hskip 80pt -24\Lambda t\sin(\Lambda t)+24(1-\cos(\Lambda t))].
\end{eqnarray}
Also, $\langle(\delta z^{1}(t))^{2}\rangle^{(0)}=\langle(\delta z^{2}(t))^{2}\rangle^{(0)}=\frac{3}{4}\langle(\delta z^{3}(t))^{2}\rangle^{(0)}$.

To estimate the magnitude of $\delta z$, we assume that the cutoff frequency to be of the order of $1/z_{0}$ as the interferometer would not be sensitive to wavelengths much shorter than its length. Suppose the time duration of the measurement $t$ is also of the order of $z_{0}$. Then one has
\begin{eqnarray}
\sqrt{(\delta z)^{2}}\sim \kappa\label{magvac}
\end{eqnarray}
which is of the order of Planck length $l_{Pl}\sim 10^{-35}$m. This is of course beyond the limit of all the present experiments.

\subsection{Thermal state}

In our discussion on the noise kernel due to gravitons in a thermal state, we see that the noise kernel separates into a vacuum part and another piece which is dependent on the temperature. This is also true for the separation correlation as well as fluctuation expressions. Since we have considered the vacuum piece in the last subsection, we shall now concentrate on the temperature dependent thermal part. Hence, from Eq.~(\ref{sepflu}), the thermal part of the separation correlation is 
\begin{eqnarray}
\langle\delta z^{i}(t)\delta z^{j}(t')\rangle^{(\beta)}
&=&\left(\frac{4}{m^{2}}\right)\delta^{ik}\delta^{jl} z_{0}^{m} z_{0}^{n}\int_{0}^{t}dt''\int_{0}^{t'}dt'''(t-t'')(t'-t''')N_{kmln}^{(\beta)}(t'',t'''),\nonumber\\
\label{thermalflu}
\end{eqnarray}
where $N_{kmln}^{(\beta)}(t,t')$ is the thermal part of the noise kernel given in Eq.~(\ref{theNK}). Although the integral can be done at least numerical, the analysis would be more transparent if we consider the low temperature and high temperature limits.

For low temperature the noise kernel can be expressed as a power series of $x=\pi(t''-t''')/\beta$ as in Eq.~(\ref{lowNK}). Note that $x$ is always small within the domains of integration. Hence, it is straightforward to evaluate the integrals as a power series in $(t-t')$.
\begin{eqnarray}
&&\langle\delta z^{i}(t)\delta z^{j}(t')\rangle^{(\beta)}\nonumber\\
&=&\left(\frac{4\pi^{4}\kappa^{2}}{945\beta^{6}}\right)\left(3\delta^{ij}\delta_{kl}+{\delta^{i}}_{k}{\delta^{j}}_{l}\right) z_{0}^{k} z_{0}^{l}\bigg[\left(\frac{t^{4}}{4}-\frac{7\pi^{2}t^{6}}{120\beta^{2}}+\frac{7\pi^{4}t^{8}}{660\beta^{4}}+\cdots\right)\nonumber\\
&&\ \ \ \ \ -\left(\frac{t^{3}}{2}-\frac{7\pi^{2}t^{5}}{40\beta^{2}}+\frac{7\pi^{4}t^{7}}{165\beta^{4}}+\cdots\right)(t-t')+\cdots\bigg].
\end{eqnarray}
Hence, the separation fluctuation is obtained by taking the limit $t\rightarrow t'$. Again, if we take $ z_{0}^{i}=(0,0, z_{0})$,
\begin{eqnarray}
\langle(\delta z^{3}(t))^{2}\rangle^{(\beta)}
=\left(\frac{16\pi^{4}\kappa^{2} z_{0}^{2}}{945\beta^{6}}\right)\left(\frac{t^{4}}{4}-\frac{7\pi^{2}t^{6}}{120\beta^{2}}+\frac{7\pi^{4}t^{8}}{660\beta^{4}}+\cdots\right)
\end{eqnarray}
and $\langle(\delta z^{1}(t))^{2}\rangle^{(\beta)}=\langle(\delta z^{2}(t))^{2}\rangle^{(\beta)}=\frac{3}{4}\langle(\delta z^{3}(t))^{2}\rangle^{(\beta)}$.

In this low temperature expansion, with $t\sim z_{0}$ again, the leading contribution to the magnitude of $\delta z$ is
\begin{eqnarray}
\sqrt{(\delta z)^{2}}\sim \kappa\left(\frac{z_{0}}{\beta}\right)^{3}.
\end{eqnarray}
Since basically $z_{0}/\beta$ is supposed to be small in this expansion, we see that there is no enhancement of this fluctuation in the low temperature case as compared to the vacuum result in Eq.~(\ref{magvac}).

For high temperature the situation is more complicated because even though $(t''-t''')/\beta$ is large in the majority of the domain of integrations in Eq.~(\ref{thermalflu}), there are places where it is small. For example, for $t''$ between 0 and $\beta$, $x=(t''-t''')\pi/\beta$ is small for $t'''$ between 0 and $t''+\beta$ while it is indeed large for $t'''$ between $t''+\beta$ and $t'$. Therefore, in the evaluation of the double integral over $t''$ and $t'''$, one needs to identify the regions where $x$ is small or large. For regions of integration with small $x$, we use the power series expansion for the noise kernel in Eq.~(\ref{lowNK}). On the other hand, for regions with large $x$, we use the approximated form of the noise kernel  in Eq.~(\ref{highNK}) by neglecting all exponentially small terms,
\begin{eqnarray}
N_{ijkl}^{(\beta)}(t,t')&=&-\frac{256\pi^{4}\alpha^{2}}{(t-t')^{6}}[2\,\delta_{ij}\delta_{kl}-3(\delta_{ik}\delta_{jl}+\delta_{il}\delta_{jk})] +\cdots.
\end{eqnarray}
With these considerations in mind, the integrations over $t''$ and $t'''$ can be performed approximately,  giving the separation correlation at the high temperature limit as,
\begin{eqnarray}
&&\langle\delta z^{i}(t)\delta z^{j}(t')\rangle^{(\beta)}\nonumber\\
&=&\left(\frac{2\pi^{4}\kappa^{2}}{\beta^{6}}\right)\left(3\delta^{ij}\delta_{kl}+{\delta^{i}}_{k}{\delta^{j}}_{l}\right) z_{0}^{k} z_{0}^{l}\bigg[\frac{\pi^{4}\beta}{89100}\left(32t^{3}-40\beta t^{2}+5\beta^{3}+\cdots\right)\nonumber\\
&&\ \ \ \ \ \ \ \ -\frac{2\pi^{4}\beta}{155925}\left(42t^{2}-35\beta t-15\beta^{2}+\cdots\right)(t-t')+\cdots\bigg].
\end{eqnarray}
Taking the $t'\rightarrow t$ limit, and also $ z_{0}^{i}=(0,0, z_{0})$, we have the separation fluctuation,
\begin{eqnarray}
\langle(\delta z^{3}(t))^{2}\rangle^{(\beta)}
=\left(\frac{64\pi^{8}\kappa^{2} z_{0}^{2}}{22275\beta^{5}}\right)\left(t^{3}-\frac{5\beta t^{2}}{4}+\frac{5\beta^{3}}{32}+\cdots\right)\nonumber\\
\end{eqnarray}
and $\langle(\delta z^{1}(t))^{2}\rangle^{(\beta)}=\langle(\delta z^{2}(t))^{2}\rangle^{(\beta)}=\frac{3}{4}\langle(\delta z^{3}(t))^{2}\rangle^{(\beta)}$.

In the high temperature expansion, the leading contribution to the magnitude of $\delta z$ is 
\begin{eqnarray}
\sqrt{(\delta z)^{2}}\sim \kappa\left(\frac{z_{0}}{\beta}\right)^{5/2}.
\end{eqnarray}
where $z_{0}/\beta$ is supposed to be large. Hence, there is an enhancement of $T^{5/2}$ where $T$ is the temperature and the fluctuations should be much more detectable. However, this is only true for the high temperature situation, like that of the early universe, but not that of the interferometric observatories like LIGO or LISA.

\subsection{Coherent state}

For a coherent state characterized by the parameter $\tilde{\alpha}$, the corresponding noise kernel is given by a vacuum part $N_{ijkl}^{(0)}$ and a coherent state part as shown in Eq.~(\ref{coherentNK}). As in the thermal case, we shall concentrate on the part related to $\tilde{\alpha}$ other than the vacuum one. 
\begin{eqnarray}
\langle\delta z^{i}(t)\delta z^{j}(t')\rangle^{(\tilde{\alpha})}
&=&\left(\frac{2\tilde{\alpha}^{2}\kappa^{2}}{9\pi}\right)\delta^{ik}\delta^{jl} z_{0}^{m} z_{0}^{n}A_{km}A_{ln}\nonumber\\
&&\ \ \int_{0}^{t}dt''\,(t'')^{-9/2}(t-t'')G(\Lambda t'')\int_{0}^{t'}dt'''\,(t''')^{-9/2}(t'-t''')G(\Lambda t'''),\nonumber\\
\end{eqnarray}
where $A_{ij}$ is the matrix defined in Eq.~(\ref{defA}) and $G(x)$ is the function defined in Eq.~(\ref{defG}). Note that the matrix $A_{ij}$ depends on the choice of a special vector. Here it is convenient to choose this special vector to be $ z_{0}^{i}/|\vec{ z}_{0}|$. Then, we have
\begin{eqnarray}
\delta^{ik}\delta^{jl} z_{0}^{m} z_{0}^{n}A_{km}A_{ln}
&=&\delta^{ik}\delta^{jl} z_{0}^{m} z_{0}^{n}\left(\delta_{km}-3\frac{ z_{0k} z_{0m}}{|\vec{ z}_{0}|^{2}}\right)\left(\delta_{ln}-3\frac{ z_{0l} z_{0n}}{|\vec{ z}_{0}|^{2}}\right)\nonumber\\
&=&4 z_{0}^{i} z_{0}^{j}.
\end{eqnarray}
The integral involving the function $G(\Lambda t)$ can be done exactly:
\begin{eqnarray}
\int_{0}^{t}dt''(t'')^{-9/2}(t-t'')G(\Lambda t'')=\Lambda^{5/2}H(\Lambda t),
\end{eqnarray}
where the function
\begin{eqnarray}
H(x)=\frac{2}{5}-\frac{3\cos x}{2x^{2}}-\frac{\sin x}{x}+\frac{3\sqrt{2\pi}}{4x^{5/2}}\,C\left(\sqrt{\frac{2x}{\pi}}\right),
\end{eqnarray}
with $C(x)$ being the Fresnel integral we have encountered before. For small $x$,
\begin{eqnarray}
H(x)=\frac{x^{2}}{9}-\frac{x^{4}}{156}+\cdots,
\end{eqnarray}
and for large $x$,
\begin{eqnarray}
H(x)&=&\left(\frac{2}{5}+\frac{3\sqrt{2\pi}}{8x^{5/2}}+\cdots\right)-\sin x\left(\frac{1}{x}-\frac{3}{4x^{3}}+\frac{9}{16x^{5}}+\cdots\right)\nonumber\\
&&\ \ \ \ \ -\cos x\left(\frac{3}{2x^{2}}+\frac{3}{8x^{4}}+\cdots\right).
\end{eqnarray}

Finally, the coherent state part of the separation correlation can be expressed as
\begin{eqnarray}
\langle\delta z^{i}(t)\delta z^{j}(t')\rangle^{(\tilde{\alpha})}
=\left(\frac{8\tilde{\alpha}^{2}\kappa^{2}}{9\pi}\right) z_{0}^{i} z_{0}^{j}\Lambda^{5}H(\Lambda t)H(\Lambda t').
\end{eqnarray}
For the separation fluctuation, we take the limit $t'\rightarrow t$ and choose $ z_{0}^{i}=(0,0, z_{0})$,
\begin{eqnarray}
\langle(\delta z^{3}(t))^{2}\rangle^{(\tilde{\alpha})}=\frac{8\tilde{\alpha}^{2}\kappa^{2} z_{0}^{2}\Lambda^{5}H^{2}(\Lambda t)}{9\pi},
\end{eqnarray}
and $\langle(\delta z^{1}(t))^{2}\rangle^{(\tilde{\alpha})}=\langle(\delta z^{2}(t))^{2}\rangle^{(\tilde{\alpha})}=0$. We can see that for $\sqrt{(\delta z^{3})^{2}}$ there is an enhancement proportional to the coherent parameter $\tilde{\alpha}$. However, for large value of $\tilde{\alpha}$, the coherent state would resemble a classical wave, that is, the classical gravitational wave. It would therefore be difficult to discern the quantum nature of gravitons in this situation.

\subsection{Squeezed state}

Consider the squeezed state parametrized by the constant $ \zeta$. The noise kernel in this state is given in Eq.~(\ref{squeezedNK}). It consists of two terms, one is $\cosh 2 \zeta $ times the Minkowski vacuum noise kernel, and the other one is $\sinh 2 \zeta$ times an expression proportional to $F(\Lambda(t+t'))$. Hence, the corresponding separation correlation will also consist of two terms. 
\begin{eqnarray}
&&\langle\delta z^{i}(t)\delta z^{j}(t')\rangle^{( \zeta)}\nonumber\\
&=&(\cosh 2 \zeta )\langle\delta z^{i}(t)\delta z^{j}(t')\rangle^{(0)}\nonumber\\
&&\ \ -(\sinh 2 \zeta )\left(\frac{\kappa^{2}}{120\pi^{4}}\right)\delta^{ik}\delta^{jl} z_{0}^{m} z_{0}^{n}B_{kmln}\Lambda^{6}\int_{0}^{t}dt''(t-t'')\int_{0}^{t'}dt'''(t'-t''')F[\Lambda(t''+t''')].\nonumber\\
\end{eqnarray}

We now concentrate on the second term. The tensor $B_{kmln}$ is defined in Eq.~(\ref{defB}). It depends on the choice of a special vector. Here again we choose it to be $\hat{ z}_{0}$. Then it is straightforward to evaluate
\begin{eqnarray}
\delta^{ik}\delta^{jl} z_{0}^{m} z_{0}^{n}B_{kmln}=-4\delta^{ij} z_{0k} z_{0}^{k}+12\, z_{0}^{i} z_{0}^{j}.
\end{eqnarray}
The integrations over $t''$ and $t'''$ are very similar to that in the Minkowski vacuum case. Expanding the result in powers of $(t-t')$,
\begin{eqnarray}
&&\int_{0}^{t}dt''\int_{0}^{t'}dt'''\,(t-t'')(t'-t''')F[\Lambda(t''+t''')]\nonumber\\
&=&\frac{1}{4\Lambda^{6}t^{2}}[-\Lambda^{4}t^{4}+2\Lambda^{2}t^{2}(1-4\cos(\Lambda t))+4\Lambda t\sin(\Lambda t)(2+\cos(\Lambda t))\nonumber\\
&&\ \ \ \ \ \ \ \ \ \ \ -2(5-4\cos(\Lambda t)-\cos^{2}(\Lambda t))]\nonumber\\
&&\ \ +\frac{(t-t')}{4\Lambda^{6}t^{3}}[\Lambda^{4}t^{4}-4\Lambda^{3}t^{3}\sin(\Lambda t)+2\Lambda^{2}t^{2}(1-2\cos(\Lambda t)-2\cos^{2}(\Lambda t))\nonumber\\
&&\hskip 70pt+4\Lambda t\sin(\Lambda t)(2+\cos(\Lambda t))-2(5-4\cos(\Lambda t)-\cos^{2}(\Lambda t))]+\cdots.\nonumber\\
\end{eqnarray}

When we take the limit $t'\rightarrow t$, the separation fluctuation takes the form
\begin{eqnarray}
\langle\delta z^{i}(t)\delta z^{j}(t)\rangle^{( \zeta)}
&=&(\cosh 2 \zeta )\langle\delta z^{i}(t)\delta z^{j}(t)\rangle^{(0)}\nonumber\\
&&\ \ +(\sinh 2 \zeta )\left(\delta_{ij}\delta_{kl}-3\delta_{ik}\delta_{jl}\right) z_{0}^{k} z_{0}^{l}\left(\frac{\kappa^{2}}{120\pi^{4}t^{2}}\right)\nonumber\\
&&\ \ \ \ \ [-\Lambda^{4}t^{4}+2\Lambda^{2}t^{2}(1-4\cos(\Lambda t))+4\Lambda t\sin(\Lambda t)(2+\cos(\Lambda t))\nonumber\\
&&\ \ \ \ \ \ \ \ \ \ \ -2(5-4\cos(\Lambda t)-\cos^{2}(\Lambda t))].
\end{eqnarray}
This result can be further simplified by choosing $\vec{ z}_{0}=(0,0, z_{0})$. Then, also putting the expression for $\langle\delta z^{i}(t)\delta z^{j}(t)\rangle^{(0)}$ in Eq.~(\ref{vacflu}), we have
\begin{eqnarray}
\langle(\delta z^{1}(t))^{2}\rangle^{( \zeta)}
&=&\langle(\delta z^{2}(t))^{2}\rangle^{( \zeta)}\nonumber\\
&=&(\cosh 2 \zeta )\left(\frac{\kappa^{2} z_{0}^{2}}{80\pi^{2}t^{2}}\right)\nonumber\\
&&\ \ \left[\Lambda^{4}t^{4}+4\Lambda^{2}t^{2}(1+2\cos(\Lambda t))-24\Lambda t\sin(\Lambda t)+24(1-\cos(\Lambda t))\right]\nonumber\\
&&\ \ +(\sinh 2 \zeta)\left(\frac{\kappa^{2} z_{0}^{2}}{120\pi^{2}t^{2}}\right)\nonumber\\
&&\ \ \ \ \ [-\Lambda^{4}t^{4}+2\Lambda^{2}t^{2}(1-4\cos(\Lambda t))+4\Lambda t\sin(\Lambda t)(2+\cos(\Lambda t))\nonumber\\
&&\ \ \ \ \ \ \ \ \ \ \ -2(5-4\cos(\Lambda t)-\cos^{2}(\Lambda t))],
\end{eqnarray}
\begin{eqnarray}
\langle(\delta z^{3}(t))^{2}\rangle^{( \zeta)}
&=&(\cosh 2 \zeta )\left(\frac{\kappa^{2} z_{0}^{2}}{60\pi^{2}t^{2}}\right)\nonumber\\
&&\ \ \left[\Lambda^{4}t^{4}+4\Lambda^{2}t^{2}(1+2\cos(\Lambda t))-24\Lambda t\sin(\Lambda t)+24(1-\cos(\Lambda t))\right]\nonumber\\
&&\ \ -(\sinh 2 \zeta )\left(\frac{\kappa^{2} z_{0}^{2}}{60\pi^{2}t^{2}}\right)\nonumber\\
&&\ \ \ \ \ [-\Lambda^{4}t^{4}+2\Lambda^{2}t^{2}(1-4\cos(\Lambda t))+4\Lambda t\sin(\Lambda t)(2+\cos(\Lambda t))\nonumber\\
&&\ \ \ \ \ \ \ \ \ \ \ -2(5-4\cos(\Lambda t)-\cos^{2}(\Lambda t))].
\end{eqnarray}

Here we see that the magnitude $\sqrt{(\delta z)^{2}}$ of the separation fluctuations has an exponential enhancement proportional either to $\sqrt{\cosh 2\zeta}$ or $\sqrt{\sinh 2\zeta }$ or $e^{\zeta }$ in the squeezed vacuum case. As the primordial gravitons produced in the inflationary era could be in a squeezed state with large squeeze parameter $\zeta$. This enhancement makes the detection of the quantum nature of primordial gravitons much more feasible \cite{Kanno,PWZ21a,PWZ21b}.

\section{Conclusions and discussions}

%PWZ model problems and resolutions
In this paper we have considered the effects of gravitons on a skeletal interferometer represented by two masses. In  the Feynman-Vernon formalism  these effects  reside in the dissipation and noise kernels of the influence action, while the quantum noise for Gaussian gravitational systems manifest as a tensorial classical stochastic  force.  From the stochastic influence action we obtained a Langevin type equation of motion for the geodesic separation of the two masses. We solved this equation in a perturbative manner to derive the corresponding correlation functions and fluctuations. This consideration actually follows from a long line of research on the stochastic dynamics of charges and masses interacting with a quantum field and in the theory of semiclassical stochastic gravity. The recent works of PWZ \cite{PWZ21a,PWZ21b} have aroused much interest in the effects of quantum noise of gravitons on the stochastic dynamics of interferometers.  These effects, if detectable, act as  a signifier of the quantum nature of gravitons.
 
In Section V, we have estimated the detectability of the geodesic separations of two masses due to the quantum noise of gravitons by calculating the magnitudes of the separation fluctuations with gravitons in different quantum states. It is found that for the Minkowski vacuum and low temperature thermal cases, the detectability of such quantum property of gravitons is slim. Although the enhancement in the high temperature case would be significant, the enviroments of the present interferometer observatories are not of this kind. The most promising case of detection would be the one with the squeezed quantum state, as the enhancement in this case is exponential. This is of course the main message from the results of  PWZ \cite{PWZ21a}. 

Our results are basically consistent with those of PWZ \cite{PWZ21a} and Kanno et al \cite{Kanno}.   PWZ considered only a selected number of modes and polarizations of the gravitons.  In spite of that, their estimations for the detectabilities of the noise effects for various graviton qantum states are in line qualitatively with our results. Our estimations based on a more complete analysis are quantitatively better.  The authors of Ref.~\cite{Kanno}  concentrated on the equations of motion rather than the effective action to derive the noise kernel. They only treated the Minkowski and the squeezed vacua, with results in agreement with ours in those cases.

Since cosmological particle creation amounts to squeezing \cite{GS90},  and as long as the squeeze parameter is sufficiently large, the separation fluctuations could be enhanced. Thus it seems to us that primordial gravitons produced in the early universe would be the most likely source for the detection of graviton noise.   Deep space experiments with long baseline interferometers (e.g., \cite{DSQL}) could provide a good platform for these quests.   The theoretical underpinning of this kind of experiments is too important to be sidelined,  namely,  the possibility for verifying the quantum nature of gravitons in perturabative quantum gravity, especially when they are placed alongside with quantum information characteristics such as gravitational decoherence (see, e.g., \cite{GravDecRev,GravDecTO}) mentioned in Sec. II.   Whether present day technologies can reveal these features depends a lot on how strongly squeezed the existent primordial gravitons are,  and the  advancement of interferometry technologies which LIGO's experience and LISA's designs would undoubtedly play a pivotal role.  These are exciting and worthy directions for further investigations into these theoretical issues and experimental possibilities.\\

\noindent {\bf Acknowledgment} HTC is supported in part by the Ministry of Science and
Technology, Taiwan, ROC, under the Grants MOST109-2112-M-032-007 and
MOST110-2112-M-032-009.

\appendix

\section{Polarization tensors}

\subsection{Polarization vectors}

To construct the polarization tensors for the gravitons, we start with the polarization vectors $\epsilon^{(1)}_{\mu}$ and $\epsilon^{(2)}_{\mu}$. Let $\vec{k}$ be the wave vector of the particle. The spatial polarization vectors $\hat{\epsilon}^{(1)}$ and $\hat{\epsilon}^{(2)}$ are orthogonal to the direction of propagation $\hat{k}=\vec{k}/|\vec{k}|$. Suppose we choose a constant unit vector $\hat{ u}_{0}=\vec{ u}_{0}/|\vec{ u}_{0}|$. Then $\hat{\epsilon}^{(1)}$ and $\hat{\epsilon}^{(2)}$ can be defined as follows. 
\begin{eqnarray}
\hat{\epsilon}^{(1)}=\hat{\epsilon}^{(2)}\times\hat{k}\ \ \ ;\ \ \ \hat{\epsilon}^{(2)}=\frac{\hat{ u}_{0}\times\hat{k}}{|\hat{ u}_{0}\times\hat{k}|}
\end{eqnarray}
where $\hat{\epsilon}^{(1)}$, $\hat{\epsilon}^{(2)}$, and $\hat{k}$ form a orthonormal basis. Therefore, the choice of this orthonormal basis depends on this particular unit vector $\hat{ u}_{0}$ which usually corresponds to some special vector present in the problem one is considering. 
For example, if one chooses $\hat{ u}_{0}=(0,0,1)$, 
\begin{eqnarray}
\hat{\epsilon}^{(1)}&=&\left(\frac{\hat{k}_{1}\hat{k}_{3}}{\sqrt{\hat{k}_{1}^{2}+\hat{k}_{2}^{2}}},\frac{\hat{k}_{2}\hat{k}_{3}}{\sqrt{\hat{k}_{1}^{2}+\hat{k}_{2}^{2}}},-\sqrt{\hat{k}_{1}^{2}+\hat{k}_{2}^{2}}\right)\nonumber\\
\hat{\epsilon}^{(2)}&=&\left(-\frac{\hat{k}_{2}}{\sqrt{\hat{k}_{1}^{2}+\hat{k}_{2}^{2}}},\frac{\hat{k}_{1}}{\sqrt{\hat{k}_{1}^{2}+\hat{k}_{2}^{2}}},0\right)
\end{eqnarray}
The 4-vector form of the polarization vectors can be written as
\begin{eqnarray}
\epsilon^{(1)}_{\mu}(\vec{k})=(0,\hat{\epsilon}^{(1)})\ \ \ ;\ \ \ \epsilon^{(2)}_{\mu}(\vec{k})=(0,\hat{\epsilon}^{(2)})\ \ \ ;\ \ \ k_{\mu}=(0,\hat{k})
\end{eqnarray}
While the circular polarization vectors are 
\begin{eqnarray}
\epsilon^{(R,L)}_{\mu}(\vec{k})=\frac{1}{\sqrt{2}}[\epsilon^{(1)}_{\mu}(\vec{k})\pm i\epsilon^{(2)}_{\mu}(\vec{k})]
\end{eqnarray}
and their complex conjugations
\begin{eqnarray}
{\epsilon^{(R,L)}_{\mu}}^{*}(\vec{k})=\epsilon^{(L,R)}_{\mu}(\vec{k})
\end{eqnarray}

In the evaluation of the noise kernels in Sec.~\ref{noisekernel}, the polarization sums of both polarization vectors and tensors are needed. Moreover, various combinations of them are integrated over solid angles. Below we list some of the formulae involved in the calculation. First, we consider the polarization sum
\begin{eqnarray}
\sum_{s=R,L}\epsilon_{i}^{(s)}(\vec{k}){\epsilon_{j}^{(s)}}^{*}(\vec{k})=P_{ij}
\end{eqnarray}
where the projection operator 
\begin{eqnarray}
P_{ij}=\delta_{ij}-\hat{k}_{i}\hat{k}_{j}
\end{eqnarray}
Without the complex conjugation, the polarization sum is more complicated.
\begin{eqnarray}
\sum_{s=R,L}\epsilon_{i}^{(s)}(\vec{k})\epsilon_{j}^{(s)}(\vec{k})=-P_{ij}+2[1-(\hat{ u}_{0}\cdot\hat{k})^{2}]^{-1}[(\hat{ u}_{0})_{i}-\hat{k}_{i}(\hat{ u}_{0}\cdot\hat{k})][(\hat{ u}_{0})_{i}-\hat{k}_{i}(\hat{ u}_{0}\cdot\hat{k})]
\end{eqnarray}
Note that this sum depends on the special unit vector $\hat{ u}_{0}$.

Next, we look at the integration over solid angles. The commonly used ones are
\begin{eqnarray}
\int\,d\Omega\,\hat{k}_{i}\hat{k}_{j}&=&\frac{4\pi}{3}\delta_{ij}\\
\int\,d\Omega\,\hat{k}_{i}\hat{k}_{j}\hat{k}_{k}\hat{k}_{l}&=&\frac{4\pi}{15}(\delta_{ij}\delta_{kl}+\delta_{ik}\delta_{jl}+\delta_{il}\delta_{jk})\\
\int\,d\Omega\,P_{ij}&=&\frac{8\pi}{3}\delta_{ij}\\
\int\,d\Omega\,P_{ij}P_{kl}&=&\frac{8\pi}{5}\delta_{ij}\delta_{kl}+\frac{4\pi}{15}(\delta_{ik}\delta_{jl}+\delta_{il}\delta_{jk})\\
\end{eqnarray}
For the polarization vector,
\begin{eqnarray}
\int\,d\Omega\,\epsilon^{(R,L)}&=&-\frac{\pi^{2}}{\sqrt{2}}\hat{ u}_{0}.
\end{eqnarray}
For two or four polarization vectors,
\begin{eqnarray}
\int\,d\Omega\,\epsilon^{(R)}_{i}\epsilon^{(R)}_{j}&=&\int\,d\Omega\,\epsilon^{(L)}_{i}\epsilon^{(L)}_{j}=-\frac{2\pi}{3}A_{ij},\\
\int\,d\Omega\,\epsilon^{(R)}_{i}\epsilon^{(L)}_{j}&=&\frac{4\pi}{3}\delta_{ij},\\
\int\,d\Omega\,\epsilon^{(R)}_{i}\epsilon^{(R)}_{j}\epsilon^{(R)}_{k}\epsilon^{(R)}_{l}&=&\int\,d\Omega\,\epsilon^{(L)}_{i}\epsilon^{(L)}_{j}\epsilon^{(L)}_{k}\epsilon^{(L)}_{l}\nonumber\\
&=&
\frac{\pi}{15}B_{ijkl}\\
\int\,d\Omega\,\epsilon^{(R)}_{i}\epsilon^{(R)}_{j}\epsilon^{(L)}_{k}\epsilon^{(L)}_{l}&=&-\frac{2\pi}{15}\left[2\,\delta_{ij}\delta_{kl}-3\,(\delta_{ik}\delta_{jl}+\delta_{il}\delta_{jk})\right],
\end{eqnarray}
where
\begin{eqnarray}
A_{ij}&=&\delta_{ij}-3(\hat{ u}_{0})_{i}(\hat{ u}_{0})_{j},\label{defA}\\
B_{ijkl}&=&(\delta_{ij}\delta_{kl}+\delta_{ik}\delta_{jl}+\delta_{il}\delta_{jk})-5\,[\delta_{ij}(\hat{ u}_{0})_{k}(\hat{ u}_{0})_{l}+\delta_{ik}(\hat{ u}_{0})_{j}(\hat{ u}_{0})_{l}\nonumber\\
&&\ \ \ \ \ +\delta_{il}(\hat{ u}_{0})_{j}(\hat{ u}_{0})_{k}+\delta_{jk}(\hat{ u}_{0})_{i}(\hat{ u}_{0})_{l}+\delta_{jl}(\hat{ u}_{0})_{i}(\hat{ u}_{0})_{k}+\delta_{kl}(\hat{ u}_{0})_{i}(\hat{ u}_{0})_{j}]\nonumber\\
&&\ \ \ \ \ +35\,(\hat{ u}_{0})_{i}(\hat{ u}_{0})_{j}(\hat{ u}_{0})_{k}(\hat{ u}_{0})_{l},\label{defB}
\end{eqnarray}
while the integrations of $\epsilon^{(R)}_{i}\epsilon^{(R)}_{j}\epsilon^{(R)}_{k}\epsilon^{(L)}_{l}$ and $\epsilon^{(R)}_{i}\epsilon^{(L)}_{j}\epsilon^{(L)}_{k}\epsilon^{(L)}_{l}$ both vanish.
These formulae will be useful when we consider integrations involving the polarization tensors.

\subsection{Polarization tensors}

In terms of the polarization vectors, the two graviton polarization tensors can be defined as
\begin{eqnarray}
\epsilon_{ij}^{+}(\vec{k})&=&\hat{\epsilon}^{(1)}_{i}(\vec{k})\hat{\epsilon}^{(1)}_{j}(\vec{k})-\hat{\epsilon}_{i}^{(2)}(\vec{k})\hat{\epsilon}_{j}^{(2)}(\vec{k})\nonumber\\
\hat{\epsilon}_{ij}^{\times}(\vec{k})&=&\hat{\epsilon}^{(1)}_{i}(\vec{k})\hat{\epsilon}_{j}^{(2)}(\vec{k})+\hat{\epsilon}_{i}^{(2)}(\vec{k})\hat{\epsilon}_{j}^{(1)}(\vec{k}).
\end{eqnarray}
One can also define the circular polarizations as
\begin{eqnarray}
\epsilon_{ij}^{(R,L)}&=&\frac{1}{\sqrt{2}}(\epsilon_{ij}^{(+)}\pm i\epsilon_{ij}^{(\times)})\nonumber\\
&=&\sqrt{2}\,\epsilon_{i}^{(R,L)}\epsilon_{j}^{(R,L)}.
\end{eqnarray}
With $\epsilon^{(R,L)}\cdot\epsilon^{{(R,L)}^{*}}=1$, we have
\begin{eqnarray}
\epsilon^{(R,L)}_{ij}\epsilon^{{(R,L)}^{*}ij}=2.
\end{eqnarray}

The sum over polarizations goes as
\begin{eqnarray}
\sum_{s=R,L}\epsilon_{ij}^{(s)}(\vec{k})\epsilon_{kl}^{(s)^{*}}=P_{ijkl}(\vec{k}),\label{polsum}
\end{eqnarray}
where the projection operator
\begin{eqnarray}
P_{ijkl}(\vec{k})=P_{ik}P_{jl}+P_{il}P_{jk}-P_{ij}P_{kl}.
\end{eqnarray}

Using the results for the integrations over solid angles for the combinations of polarization vectors, we can obtain the following formulae for the corresponding polarization tensors.
\begin{eqnarray}
\int\,d\Omega\,\epsilon_{ij}^{(R,L)}(\vec{k})&=&-\frac{2\sqrt{2}\pi}{3}A_{ij},\\
\int\,d\Omega\,\epsilon_{ij}^{(R)}\epsilon_{kl}^{(R)}
&=&\int\,d\Omega\,\epsilon_{ij}^{(L)}\epsilon_{kl}^{(L)}\nonumber\\
&=&\left(\frac{2\pi}{15}\right)B_{ijkl},\\
\int\,d\Omega\,\epsilon_{ij}^{(R)}\epsilon_{kl}^{(L)}
&=&-\left(\frac{4\pi}{15}\right)[2\,\delta_{ij}\delta_{kl}-3\,(\delta_{ik}\delta_{jl}+\delta_{il}\delta_{jk})],
\end{eqnarray}

For polarization sums,
\begin{eqnarray}
\int\,d\Omega\,\sum_{s}\epsilon_{ij}^{(s)}\epsilon_{kl}^{(s)^{*}}
&=&\int\,d\Omega\,P_{ijkl}\nonumber\\
&=&-\left(\frac{8\pi}{15}\right)[2\,\delta_{ij}\delta_{kl}-3\,(\delta_{ik}\delta_{jl}+\delta_{il}\delta_{jk})],\label{intpijkl}\\
\int\,d\Omega\,\sum_{s}\epsilon_{ij}^{(s)}\epsilon_{kl}^{(s)}
&=&\left(\frac{4\pi}{15}\right)B_{ijkl}\label{intsum}.
\end{eqnarray}

%%%%%%%%%%%%%%%%%%%%%%%%%%%%%%%%%
%
% References
%

 \end{document}